\documentclass[prd,nofootinbib,twocolumn,superscriptaddress,showpacs]{revtex4}

\usepackage[usenames]{color}
\usepackage{amsfonts}
\usepackage{amsmath}
\usepackage{amssymb}
\usepackage{bm}
\usepackage{dcolumn}
\usepackage{epsfig}
\usepackage{graphicx}
\usepackage{graphics}
\usepackage[latin1]{inputenc}
\usepackage{latexsym}
\usepackage{rotating}
\usepackage{hyperref}
\usepackage{multirow}
\usepackage[caption=false]{subfig}
\usepackage[usenames,dvipsnames]{xcolor}
\usepackage[normalem]{ulem} % To get strikethrough (\sout)

\newcommand{\<}{\begin{equation}}
\newcommand{\?}{\end{equation}}

\newcommand{\N}{\mathbb{N}}

\newcommand{\Q}{\mathbb{Q}}

\newcommand{\cG}{\mathcal{G}}

\newcommand{\cR}{\mathcal{R}}

\newcommand{\cS}{\mathcal{S}}
\newcommand{\cT}{\mathcal{T}}
\newcommand{\cU}{\mathcal{U}}
\newcommand{\cV}{\mathcal{V}}
\newcommand{\cX}{\mathcal{X}}

\DeclareMathOperator{\eulerlog}{eulerlog}

\newcommand{\od}[2]{\left[ #1 \right]_{ #2 }}

\newcommand{\ri}{\mathrm{i}}
\newcommand{\amp}{\;\text{\&}\;}

\newcommand{\bea}{\begin{eqnarray}}
\newcommand{\eea}{\end{eqnarray}}
\newcommand{\bsube}{\begin{subequations}}
\newcommand{\esube}{\end{subequations}}
\newcommand{\beq}{\begin{equation}}
\newcommand{\be}{\begin{equation}}
\newcommand{\eeq}{\end{equation}}
\newcommand{\ee}{\end{equation}}

\begin{document}

\title{Experimental mathematics meets gravitational self-force}

\author{Nathan~K.~Johnson-McDaniel}

 \affiliation{Theoretisch-Physikalisches Institut,
  Friedrich-Schiller-Universit{\"a}t,
  Max-Wien-Platz 1,
  07743 Jena, Germany}
  
  \affiliation{International Centre for Theoretical Sciences,
  Tata Institute of Fundamental Research,
  Bengaluru 560012, India
  }
  
  \author{Abhay~G.~Shah}
  
  \affiliation{Mathematical Sciences, University of Southampton, Southampton SO17 1BJ, United Kingdom}

  \author{Bernard~F.~Whiting}
  
  \affiliation{Institute for Fundamental Theory, Department of Physics, University of Florida, Gainesville, Florida 32611, USA}
  
\date{\today}

%--------------------------
\begin{abstract}

It is now possible to compute linear in mass-ratio terms in the post-Newtonian (PN) expansion for compact binaries to very high orders using linear black hole perturbation theory applied to various invariants. For instance, a computation of the redshift invariant of a point particle in a circular orbit about a black hole in linear perturbation theory gives the linear-in-mass-ratio portion of the binding energy of a circular binary with arbitrary mass ratio. This binding energy, in turn, encodes the system's conservative dynamics. We give a method for extracting the analytic forms of these post-Newtonian coefficients from high-accuracy numerical data using experimental mathematics techniques, notably an integer relation algorithm. Such methods should be particularly
important when the calculations progress to the considerably more difficult case of perturbations of the Kerr metric. As an example, we apply this method to the redshift invariant in Schwarzschild. Here we obtain analytic coefficients to $12.5$PN, and higher-order terms in mixed analytic-numerical form to $21.5$PN, including analytic forms for the complete $13.5$PN coefficient, and all the logarithmic terms at $13$PN. We have computed the individual
modes to over $5000$ digits, of which we use at most $1240$ in the present calculation. At these high orders, an individual coefficient can have over $30$ terms, including a wide variety of transcendental numbers, when written out in full. We are still able to obtain analytic forms for such coefficients from the numerical data through a careful study of the structure of the expansion. The structure we find also allows us to predict certain ``leading logarithm''-type contributions to all orders. The additional terms in the expansion we obtain improve the accuracy of the PN series for the redshift observable, even in the very strong-field regime inside the innermost stable circular orbit, particularly when combined with exponential resummation.
\end{abstract}
%--------------------------


\maketitle

%%%%%%%%%%%%%%%%%
\section{Introduction and Summary}
%%%%%%%%%%%%%%%%%

Coalescing compact binaries are a promising source of gravitational waves, and ground-based gravitational wave interferometers will start
operating at sensitivities at which detections can reasonably be expected as early as later this  year. In order to successfully detect these faint signals in the detector's noise---and, more importantly, to be able to infer the properties of the system from the detected signal---it is necessary to have highly accurate templates that model the gravitational waves from the inspiralling binaries. Thus, for more than a decade now, different approaches have been developed to model relativistic binary systems. The oldest one of these, the post-Newtonian (PN) framework, can model such systems when the two bodies are far from one another, so their velocities are relatively slow (see~\cite{BlanchetLRR} for a review of these methods and results). Numerical relativity, on the other hand, is able to model comparable mass ratio binaries in the strong gravitational field regime, but has difficulties with large mass
ratios, large separations, and very long waveforms (but see~\cite{NZLC, LZ-outerlimits, Szilagyi_etal} for recent advances). Another approach is gravitational self-force theory, which models binaries with extreme mass-ratios, where one has a small body that is about a million times lighter than the central super-massive black hole into which it is inspiralling~\cite{Thornburg-GWN,Barack-review,Amaro-Seoaneetal}.

A more recent approach, effective-one-body (EOB) theory, maps the binary's motion to that of a particle moving in an effective metric, generalizing the Newtonian reduced-mass treatment of the two-body problem~\cite{DN-review,Damour-review}. This theory encompasses information from the former three approaches to calibrate the parameters that go into the theory, which allows it to model a binary system of any given mass-ratio.  Of particular interest is the overlap region between the self-force and PN formalisms. Invariant quantities calculated in this region are used to calibrate the EOB parameters. One of those quantities calculated in self-force theory is Detweiler's redshift invariant, $\Delta U$, the linear-in-mass-ratio correction to the time component of the 4-velocity of the light compact object~\cite{Detweiler}. The PN coefficients of $\Delta U$ are directly related to those of the linear-in-mass-ratio portion of the binding energy and angular momentum of the binary, as well as to the radial potential that is fundamental to the EOB formalism,
as was demonstrated in~\cite{bd_14,djs_15,bblt_16,ltbw,ltbb_13,ABDS}.

Computation of the PN coefficients of $\Delta U$ started with Detweiler's original paper~\cite{Detweiler} (to $2$PN; $n$PN corresponds to an accuracy of $v^{2n}$, where $v$ is the orbital velocity of the small body), and continued analytically through $3$PN in~\cite{BDLtW1}, using standard post-Newtonian methods, with terms through $5$PN obtained from a numerical matching in~\cite{BDLtW2} (where the logarithmic terms were obtained analytically). In~\cite{BD1}, Bini and Damour calculated the nonlogarithmic portion $4$PN coefficient analytically by using analytical solutions of the Regge-Wheeler-Zerilli equations; the logarithmic term had already been
computed by Damour in~\cite{Damour10}. Shah, Friedman, and Whiting~\cite{SFW} (hereafter referred to as SFW) calculated higher order PN coefficients, up to $10.5$PN order, by calculating high-precision numerical values in a modified radiation gauge at very large radii and fitting them to a PN series to extract the coefficients. They found that half-integer terms started at $5.5$PN, which they verified analytically (and was also verified using standard PN methods in~\cite{BFW1,BFW2}). SFW were also able to infer analytic forms for certain not-too-complicated coefficients from their numerical data. Concurrently, Bini and Damour calculated the coefficients analytically to $6$PN order in~\cite{BD2}. Analytical calculations since then~\cite{BD3,BD6,KOW}, all using the Regge-Wheeler-Zerilli gauge and the results from~\cite{MST2}, have been pushed to find much higher order PN coefficients. We have compared our results to $20.5$PN with the concurrent calculation  by Kavanagh, Ottewill, and Wardell~\cite{KOW}, who have streamlined the Bini-Damour method, and found
complete agreement. (Note that Bini and Damour's $9.5$PN result~\cite{BD6} appeared while we were finishing checking higher-order terms in this work.) 

Apart from $\Delta U$, high-order PN coefficients of other invariants from which the EOB formalism can benefit have also been calculated: These are the linear-in-mass ratio conservative corrections to the spin-precession angle~\cite{Dolanetal-spin} and the quadrupolar~\cite{Dolanetal-tidal} and octupolar~\cite{Nolanetal-tidal} tidal invariants (the eigenvalues of the electric- and magnetic-type tidal tensors), all which have been calculated to high PN order in \cite{BD-spin, BD-tidal, KOW, SP, Nolanetal-tidal}. Recently, Bernuzzi~\emph{et al.}~\cite{Bernuzzietal} introduced a semi-analytical tidally coupled binary neutron star model using the EOB theory where information from the PN expansion of the redshift and tidal invariants was incorporated (using the tidal results from Bini and Damour~\cite{BD-tidal}). The results of this model are in good agreement with a more recent numerical simulation in full general relativity by the Japanese school (Hotokezaka~\emph{et al.}~\cite{Hotokezakaetal}) in the case of compact neutron stars. 

It has recently been shown (see \cite{SFW,SKerr, SP}) how the overlap region between the self-force formalism and the PN approximation can be explored using very high accuracy numerical results, which make it relatively easy to extract high-order PN coefficients that are currently out of reach of standard PN calculations. The coefficients obtained using this numerical extraction method have then been checked by independent analytical calculations. The advantage of developing such high-accuracy calculations will be evident when PN coefficients are calculated for invariants in Kerr spacetime, where purely analytical calculations will likely be extremely difficult. The techniques developed in this paper can then be generalized to calculate the analytical form of the numerical coefficients for various invariants in Kerr, and, eventually, to calculate the quadratic-in-mass-ratio terms using second-order self-force results (see Sec.~4.3.3 in~\cite{Amaro-Seoaneetal} for a brief overview of progress in second-order self-force calculations). Additionally, one obtains insight into the structure of the PN expansion
from these high-order computations, particularly in comparing the terms one can predict using a simplification in $\Delta U$ with similar terms in the energy flux at infinity~\cite{NKJ-M_tamingPN}. 

We shall now outline our method and compare it to previous work. Shah, Friedman, and Whiting (SFW)~\cite{SFW} worked solely on the expansion of the full $\Delta U$ and obtained analytic terms for the
simplest coefficients, which are purely rational, or a rational times $\pi$, where they could easily identify the analytic form from a large enough number of digits. They also present three additional analytic expressions for more complicated  higher-order terms (in the note added), which were obtained by the first author of this paper using an integer relation algorithm. However, the accuracy of the expressions in SFW was insufficient to obtain analytic forms for any terms beyond $10.5$PN order.

The methods we use here are similar to those used to obtain the more complicated coefficients given in SFW (and the analytic coefficients given in~\cite{SKerr,SP}), in that we also use an integer relation algorithm, but the present application is more effective at obtaining higher-order terms, since we primarily work with the individual modes of $\Delta U$ [either retarded $(\ell, m)$ modes, or renormalized $\ell$ modes], where the structure of the expansion is simpler, and one can obtain analytic forms at a given order with fewer digits. Indeed, one can often predict some---and in certain cases even all---of the entire analytic form at higher orders from lower-order coefficients. This simplification of the structure when considering the individual $(\ell, m)$ modes was also seen in the expansion of the energy flux at infinity of a point particle in a circular orbit around a Schwarzschild black hole~\cite{NKJ-M_tamingPN}. Additionally, the overall structure of the expansion of the retarded $(\ell, m)$ modes of $\Delta U$ is also similar to that of the energy flux at infinity (calculated to $22$PN by Fujita~\cite{Fujita22PN}, with the structure studied in~\cite{NKJ-M_tamingPN}), and we are able to use this to help determine which transcendentals to include in the vector to which we applied the integer relation algorithm.

We also use the integer relation algorithm in a more fundamental way in the current work, preferring for most of our work to find analytic expressions for the terms order-by-order and then subtract them off to obtain the numerical values for higher-order terms to higher accuracy. (Note that we found that for certain of the more complicated terms at higher orders it was necessary to first obtain analytic forms for some of the simpler coefficients at even higher orders in order to obtain the more complicated terms to sufficient accuracy to be able to determine an analytic form.) This method should be contrasted with the more usual method of finding numerical values for all terms to some accuracy using a fit, then finding analytic forms for some coefficients, using these to improve the accuracy of the fit, and iterating. This fitting method was used in SFW and in Nickel's similar computation of high-order terms in the expansion of the ground state energy of $H_2^+$ in powers of the distance~\cite{Nickel}.\footnote{This expansion of the ground state energy of $H_2^+$ has a similar structure to an individual mode of $\Delta U$ (though it is simpler), and can be computed using functional series techniques similar to those we employ here, as discussed in~\cite{LeaverJMP}.} We also used this fitting method on the full $\Delta U$ to obtain even higher-order terms than we were able to obtain using the first method, though these terms were all obtained only in mixed numerical-analytic form.

These integer relation algorithms, notably the PSLQ algorithm~\cite{FB,FBA}, are a prominent tool in modern experimental mathematics. (See also~\cite{Straub,CSV} for further intuition into the PSLQ algorithm and~\cite{Bailey} for a review of some remarkable results obtained using integer relation algorithms. Additionally, see~\cite{BB_AMS, BBBZ,BB_AMS2} for some general reviews of the methods, philosophy, and results of modern experimental mathematics.) The PSLQ algorithm returns a small vector of integers that is orthogonal to a given input vector, and thus can be used to identify the analytic form of numbers from a high-accuracy decimal expansion, which is the task for which we use it here, employing the implementation in the FindIntegerNullVector function in {\sc{Mathematica}} (first available in version 8).

Here we only need to identify numbers that are linear combinations of transcendentals with rational coefficients, which is one of the simplest cases to which one can imagine applying an integer relation algorithm. Nevertheless, there are enough transcendentals at higher orders, with complicated enough rational coefficients, that we still need to compute certain individual PN coefficients to over $200$ digits, even when using a simplification we found that helps remove much of the complexity at higher orders. This necessitates computing the modes of $\Delta U$ to over $1000$ digits; we actually computed to over $5000$ digits so we could go to even higher orders, where we currently only obtain certain coefficients analytically.
 Some other such high-accuracy computations in mathematical physics, including further applications of PSLQ, are discussed in~\cite{BB_Mathematics}. Additionally, Nickel~\cite{Nickel} also uses PSLQ to obtain analytic forms of high-order coefficients of a similar series for the ground-state energy of $H_2^+$.

We also note, following Bini and Damour~\cite{BD2}, that while one has to sum over all spherical harmonic $(\ell, m)$ modes to obtain $\Delta U$ to a given PN order (compared to, e.g., the energy flux, where one only has to sum a finite number of modes to obtain the expansion to a given PN order), this infinite sum is only necessary to obtain the nonlogarithmic integer-order PN terms. All the other terms in the PN expansion of $\Delta U$ come from a finite sum over modes. 

It also turns out that the expression for the PN coefficient of a given renormalized $\ell$ mode (at high $\ell$, where it is purely rational) is simple enough that one can infer it from the numerical values of fewer than $100$ $\ell$-modes, at the PN orders at which we are working, so we can obtain these general expressions and then perform the infinite $\ell$-sum analytically, allowing us to calculate analytic forms for the nonlogarithmic integer-order PN coefficients of $\Delta U$ without performing the $\ell$-sum numerically. This is fortunate, since performing the infinite $\ell$-sum numerically to such high accuracies would be prohibitively expensive, computationally, due to
the necessity of calculating many $\ell$ modes. We obtained the full expansion up to $12.5$PN this way (including reproducing all the known analytic terms ``from scratch'').

We can even obtain fairly complicated forms of high-order terms (though not any complete PN terms) using the predictions of the simplification, PSLQ, and a reasonably (but not excessively) accurate calculation of the full $\Delta U$. In particular, we performed a calculation of the full $\Delta U$ to ``merely'' $\sim 600$ digits at somewhat smaller radii ($10^{18}M$ to $9\times 10^{33}M$, where $M$ is the mass of the central object) to check the values we obtained using the data calculated to more than $5000$ digits for radii from $10^{50}M$ to $10^{70}M$ (where we used at most $1240$ digits and $15$ radii to obtain those results).\footnote{These calculations
were computationally not exceptionally expensive, requiring $\sim 45$~hours per radius on $2$ processors at $\sim 600$ digits, and $\sim 10$~hours per radius on $16$ processors at $\sim 5000$ digits.} Using the results of this calculation, we were able to obtain accurate values up to $21.5$PN, including analytic forms for $48$ coefficients containing as many as $27$ terms (most coefficients had far fewer terms), starting from the full $\Delta U$, though most of these terms were predicted by the simplification: We only had to use PSLQ to obtain at most $4$ terms. Here we also used the analytic forms we obtained to iteratively improve the PN coefficients, increasing the accuracy of terms we had already obtained, in addition to obtaining even higher-order terms.

The plan of the paper is as follows. We first briefly review the relevant portions of the self-force calculation in Sec.~\ref{sec:sf}, and then discuss the method we use to obtain the PN coefficients of the individual modes of $\Delta U$ (including a simplification of the modes, and consistency checks) in Sec.~\ref{sec:modes}. We then give the terms in the full $\Delta U$ that are predicted to all orders by the simplification of the modes in Sec.~\ref{sec:simp}, and discuss how we compute the infinite sum over the modes of $\Delta U$ to obtain the final results for the PN coefficients of $\Delta U$ (and our independent check of these results) in Sec.~\ref{sec:sum}. We discuss convergence of the series in Sec.~\ref{sec:conv} and conclude in Sec.~\ref{sec:concl}. In the Appendix, we give some discussion of how one can obtain certain parts of the simplifications of the modes of $\Delta U$ from inspection of the method we use to calculate it. We use geometrized units throughout (setting the speed of light and Newton's gravitational constant both to unity, i.e., $G = c = 1$).

%%%%%%%%%%%%%%%%%
\section{Self-force calculation}
\label{sec:sf}
%%%%%%%%%%%%%%%%%

Here we give the basics of the method we use to calculate $\Delta U$ (and its precise definition)---see~\cite{sf2,sf3,sf4,SFW} for further details.
We calculate $\Delta U$ in a modified radiation gauge, where $\ell \ge 2$ modes are calculated in an outgoing radiation gauge (with $h_{\alpha\beta}n^\alpha=0$ and $h=0$, where $n^\alpha$ is the ingoing null vector and $h_{\alpha\beta}$ and $h$ are the metric perturbation and its trace, respectively) and the lower ones ($\ell=0,1$) are calculated in the asymptotically flat Schwarzschild gauge. The setup is as follows: A particle of mass $\mathfrak{m}$ is orbiting a Schwarzschild black hole of mass $M$ in a circular orbit of radius $r=r_0$ in Schwarzschild coordinates $(t,r,\theta,\phi)$. The particle's four-velocity, $u^\alpha$, is given by
\begin{align}
u^\alpha = u^t t^\alpha + u^\phi \phi^\alpha,
\end{align}
where $t^\alpha$ and $\phi^\alpha$ are the time-like and rotational Killing vectors of the Schwarzschild metric, respectively. The components, $u^t$ and $u^\phi$, are given by
\begin{align}
U := u^t &= \frac{1}{\sqrt{1-\frac{3M}{r_0}}}, \\ %\textrm{and}
u^\phi &= u^t \Omega, \, \textrm{with}\\
\Omega &= \sqrt{\frac{M}{r_0^3}}.
\end{align}
We follow the Chrzanowski-Cohen-Kegeles-Wald formalism (outlined in \cite{sf2})
of extracting the metric perturbation from the perturbed spin-2 Weyl scalar $\psi_0$ as follows. We first solve the spin-2 separable Teukolsky equation, whose retarded solution, $\psi_0$ (the superscript ``ret" is omitted here), is given by
\begin{align}
\psi_0(x) = \psi_0^{(0)}+\psi_0^{(1)}+\psi_0^{(2)},
\label{eq:psiGT}
\end{align}
with
\begin{widetext}
\bsube\label{eq:psi0}
\bea
\psi_0^{(0)} &=& 4\pi {\mathfrak m} u^t \frac{\Delta_0^2}{r_0^2}\sum_{\ell m}A_{\ell m}[(\ell-1)\ell(\ell+1)(\ell+2)]^{1/2}R_{\rm H}(r_<)R_\infty(r_>){}_2Y_{\ell m}(\theta,\phi)\bar{Y}_{\ell m}\left(\frac{\pi}{2},\Omega t\right), \\
\psi_0^{(1)} &=& 8\pi \ri{\mathfrak m} \Omega u^t \Delta_0 \sum_{\ell m} A_{\ell m}[(\ell-1)(\ell+2)]^{1/2}
	{}_2Y_{\ell m}(\theta,\phi){}_1\bar{Y}_{\ell m}\left(\frac{\pi}{2},\Omega t\right) \times  \nonumber\\
& & \quad \Bigl\{[\ri m\Omega r_0^2 + 2 r_0]R_{\rm H}(r_<)R_\infty(r_>) 
	+ \Delta_0[R_{\rm H}'(r_0)R_\infty(r)\theta(r-r_0) 	 + R_{\rm H}(r)R_\infty'(r_0)\theta(r_0-r)]\Bigr\},\\
\psi_0^{(2)} &=& -4\pi {\mathfrak m}\Omega^2 u^t \sum_{\ell m} A_{\ell m}
	{}_2Y_{\ell m}(\theta,\phi){}_2\bar{Y}_{\ell m}\left(\frac{\pi}{2},\Omega t\right) \times 
\nonumber\\ 
& & \biggl\{[30r_0^4 - 80Mr_0^3 + 48M^2r_0^2 - m^2\Omega^2 r_0^6 -2\Delta_0^2 - 24\Delta_0 r_0(r_0-M)+ 6\ri m\Omega r_0^4(r_0-M)]
	R_{\rm H}(r_<)R_\infty(r_>)
 \nonumber\\ 
& & \qquad  + 2(6r_0^5 - 20Mr_0^4 + 16M^2r_0^3 - 3r_0\Delta_0^2 + \ri m\Omega \Delta_0 r_0^4)[R_{\rm H}'(r_0)R_\infty(r)\theta(r-r_0) 
 + R_\infty'(r_0)R_{\rm H}(r)\theta(r_0-r)]\nonumber\\ 
& & \qquad + r_0^2\Delta_0^2[R_{\rm H}''(r_0)R_\infty(r)\theta(r-r_0) + R_\infty''(r_0)R_{\rm H}(r)\theta(r_0-r)+\textrm{W}[R_{\rm H}(r),R_\infty(r)]\delta(r-r_0)]\Biggr\},
%& \qquad +\textrm{W}\delta(x-x_0)]
\eea\esube 
\end{widetext}
where $\Delta := r^2 - 2Mr$; the function $R_H$ is the solution of the homogenous radial Teukolsky equation which is ingoing at the future event horizon and $R_\infty$ is the one that is outgoing at null infinity. Here $r_< := \textrm{min}(r,r_0)$ and $r_> := \textrm{max}(r,r_0)$.
%\nkjm{Don't we need to define $r_<$ and $r_>$? Also, I removed the ``$re$'' from the end of the third line of $\psi_0^{(2)}$, as it seemed to be anomalous: While it's present in SFW, it's not present in Eq.~(18c) in~\cite{sf3}.} 
A prime denotes a derivative with respect to the $r$-coordinate and overbars denote complex conjugation; $\delta$ and $\theta$ denote the Dirac delta distribution and Heaviside theta function, respectively. The Wronskian of these two retarded radial solutions is $\textrm{W}[R_{\rm H}(r),R_\infty(r)] = R_{\rm H} R_\infty^\prime - R_\infty R_{\rm H}^\prime $.  The quantity $A_{\ell m}$,
given by 
\be
 A_{\ell m} := \frac{1}{\Delta^3 \textrm{W}[R_{\rm H}(r),R_\infty(r)]},
\label{eqAlm}
\ee
is a constant independent of $r$, that is, $A_{\ell m}^\prime=0$.  
The functions $R_H$ and $R_\infty$ are calculated to more than $5000$ digits of accuracy using the \emph{Mano, Suzuki, and Takasugi (MST) method} given in~\cite{MST}, namely
\begin{widetext}
\begin{subequations}
\begin{align}
R_H &= e^{\ri\epsilon x}(-x)^{-2-\ri\epsilon}\sum_{n=-\infty}^{\infty}a_n F(n+\nu+1-\ri\epsilon,-n-\nu-\ri\epsilon,-1-2\ri\epsilon;x), 
\label{RH}\\
R_\infty &= e^{\ri z}z^{\nu-2} \sum_{n=-\infty}^{\infty} (-2z)^n b_n U(n+\nu+3-\ri\epsilon,2n+2\nu+2;-2\ri z).
\label{radialfunctions}
\end{align}
\end{subequations}
\end{widetext}
where $x = 1-\frac{r}{2M}$, $\epsilon = 2 M m\Omega$, $z = -\epsilon x$, $F$ is the hypergeometric function $\,_2F_1$, and $U$ is the (Tricomi) confluent hypergeometric function. 
%\nkjm{Say something here (or slightly later) about the recursion relations you used to obtain such high accuracy. You might also mention how many terms you had to sum to achieve this accuracy...} 
%
To expedite the calculation, and reach the high accuracies we require, we use various recurrence relations for $U$, and Gauss's relations for contiguous functions for $\,_2F_1$ (see, e.g.,~\cite{AS}) to write the various $n$-dependent functions and their derivatives in terms of the functions calculated for $n=0,1$.
For details regarding the derivation of $\nu$, the renormalized angular momentum, and the coefficients $a_n$ and $b_n$, please refer to \cite{MST,ST}.
The spin-weighted spherical harmonics $_sY_{\ell m}(\theta,\phi)$ are calculated analytically and are given in \cite{sf3}.

From $\psi_0$, we compute the intermediate Hertz potential, $\Psi$, from which the components of the metric perturbation are calculated. The radial parts of $\Psi$ and $\psi_0$ are related by an algebraic relation given by
\beq\label{eq:Psi}
\Psi_{\ell m} = 8 \frac{(-1)^m (\ell+2)(\ell+1)\ell(\ell-1)\bar\psi_{\ell,-m}
	+ 12\ri m M \Omega \psi_{\ell m} }{ [(\ell+2)(\ell+1)\ell(\ell-1)]^2 + 144 m^2 M^2 \Omega ^2},
\eeq
where 
\begin{subequations}
\begin{align}
\Psi &= \sum_{\ell,m}\Psi_{\ell m}(r){\,}_2Y_{\ell m}(\theta,\phi)e^{-\ri m\Omega t}, \\
\psi_0 &= \sum_{\ell,m}\psi_{\ell m}(r){\,}_2Y_{\ell m}(\theta,\phi)e^{-\ri m\Omega t}.
\end{align} 
\end{subequations}
Once we compute $\Psi$, the components of the metric perturbation along the Kinnersley tetrad are given by 
%\begin{widetext}
\<
\begin{split}
h_{\bf 11} &= \frac{r^2}{2}(\bar{\eth}^2\Psi + \eth^2\overline\Psi),  \\ 
h_{\bf 33} &= r^4\biggl[\frac{\partial_t^2 -2f\partial_t\partial_r+f^2\partial_r^2}{4} - \frac{3(r-M)}{2r^2}\partial_t\\
&\quad + \frac{f(3r-2M)}{2r^2}\partial_r + \frac{r^2-2M^2}{r^4}\biggr]\Psi, \\
h_{\bf 13} &= -\frac{r^3}{2\sqrt{2}}\left(\partial_t-f\partial_r-\frac{2}{r}\right)\bar{\eth}{\Psi},
\end{split}
\?
%\end{widetext}
where $f = \Delta/r^2$, and the angular operators $\eth$ and $\bar{\eth}$, the s-raising and -lowering operators, 
are given by
\<
\begin{split}
\eth\eta &= -\left(\partial_\theta+\ri\csc\theta\partial_\phi-s\cot\theta\right)\eta, \\
\bar{\eth}\eta&=-\left(\partial_\theta-\ri\csc\theta\partial_\phi+s\cot\theta\right)\eta,
\end{split}
\?
where $\eta$ has spin-weight $s$.

The linear-in-mass-ratio correction to the time component of the four-velocity of the particle due to its finite but small mass is then given by 
\begin{align}\label{eq:DU}
\Delta U = -U H^\textrm{ren}
\end{align}
where
\begin{align}
H^\textrm{ren} = \frac12 h_{\alpha\beta}^\textrm{ren} u^\alpha u^\beta.
\end{align}
The super-script ``ren" denotes the renormalized, singularity-free part of the metric perturbation. We refer the reader to \cite{MiSaTa,QuWa,DW,sf2,sf3,sf4,SFW} for details pertaining to the renormalization procedure.

As mentioned earlier, $\psi_0$ only provides us with the radiative part of the metric perturbation. One also has to add on the non-radiative parts associated with the change in mass and angular momentum of the Schwarzschild spacetime with the particle. These contributions are given by (Eqs.~(137) and (138) of \cite{sf4})
\begin{subequations}
\begin{align}
H_{\delta M} & = \frac{\mathfrak{m}(r_0-2M)}{r_0^{1/2}(r_0-3M)^{3/2}}, \\
H_{\delta J} &= \frac{-2M\mathfrak{m}}{r_0^{1/2}(r_0-3M)^{3/2}}.
\end{align}
\end{subequations}
The index $\delta M$ and $\delta J$ refer to the parts coming from the change in mass and angular momentum, respectively. Also note that the $\ell=1,m=\pm1$ (even) contribution to such gauge-invariant quantities, corresponding to the shift in center-of-mass of the binary $\frak{m}-M$ system, is zero.

%%%%%%%%%%%%%%%%%
\section{Obtaining analytic forms of the PN coefficients of the individual $(\ell,m)$ modes of $\Delta U$}
\label{sec:modes}
%%%%%%%%%%%%%%%%%

%-------------------------
\subsection{The PN expansion of the $(2,2)$ mode of $\Delta U$ and a way to simplify a general $(\ell, m)$ mode of $\Delta U$}
%-------------------------

We start by giving our expression for the PN expansion of $\Upsilon_{22}$, the $(2,2)$ mode of $\Delta U/U$, through $12.5$PN, as well as the simplification of the modes we have discovered, before describing our method for obtaining these results. [We give the analogous results for the other modes in the electronic Supplemental Material~\cite{J-MSW_suppl}, along with higher-order PN coefficients in the $(2,2)$ mode for which we only know analytic forms for some of the terms, and the $13.5$PN piece we do know all of.] Here we consider $\Delta U/U$ instead of just $\Delta U$ as this is the quantity that we worked with on the level of the individual modes [note that $\Delta U/U = -H^\text{ren}$; cf.\ Eq.~\eqref{eq:DU}]. We present the expansion in terms of the same dimensionless and gauge invariant radius variable used in SFW, viz., $R := (M\Omega)^{-2/3} = r_0/M$:
\begin{widetext}
\<
\label{eq:Upsilon22}
\begin{split}\nonumber
\Upsilon_{22} &=
%\frac{3}{4}\frac{1}{R} + \frac{19}{56}\frac{1}{R^2} + \frac{7403}{2016}\frac{1}{R^3} + \frac{5813077}{2587200}\frac{1}{R^4} + \left[\frac{1090372790267}{16951334400} - \frac{128}{5}\eulerlog_2(R)\right]\frac{1}{R^5}\\
%&\quad + \left[\frac{16555266905081}{305124019200} + \frac{9664}{105}\eulerlog_2(R)\right]\frac{1}{R^6} - \frac{13696}{525}\frac{\pi}{R^{6.5}} + \left[\right]
\frac{3^1}{2^2}\frac{1}{R}+\frac{19^1}{2^3 7^1}\frac{1}{R^2}+\frac{11^1 673^1}{2^5 3^2 7^1}\frac{1}{R^3}+\frac{5813077^1}{2^6 3^1 5^2 7^2 11^1}\frac{1}{R^4}+\left[\frac{19^1 727^1 78938159^1}{ 2^9 3^3 5^2 7^3 11^1 13^1}-\frac{2^7}{5^1}\eulerlog_2(R)\right]\frac{1}{R^5}\\
&\quad+ \left[\frac{29^1 71^1 733^1 3301^1 3323^1}{ 2^{10} 3^5 5^2 7^3 11^1 13^1} + \frac{2^6151^1 }{3^15^17^1}\eulerlog_2(R)\right]\frac{1}{R^6} - \frac{2^7 107^1}{3^1 5^2 7^1}\frac{\pi}{R^{6.5} }+
\biggl[-\frac{ 6217^1 51217^1 95947^1 108961^1}{ 2^{12} 3^4 5^4 7^4 11^2 13^1 17^1}\\
&\quad + \frac{2^4 6899^1}{3^3 5^1 7^2}\eulerlog_2(R)\biggr]\frac{1}{R^7}+\frac{2^6 107^1 151^1}{3^2 5^2 7^2}\frac{\pi}{R^{7.5}}+
\biggl[\frac{181399^1 10818902184042899087^1}{2^{13} 3^6 5^4 7^5 11^2 13^2 17^1 19^1}-\frac{2^8 107^1}{3^2 5^1 7^1}\pi ^2 -\frac{2^{11}}{5^1}\zeta (3)\\
&\quad -\frac{2^3 1529121911^1}{3^4 5^3 7^2 11^1}\eulerlog_2(R) + \frac{2^{10}107^1}{3^15^27^1}\eulerlog_2^2(R)\biggr]\frac{1}{R^8}+\frac{2^4 107^1 6899^1}{ 3^4 5^2 7^3}\frac{\pi}{R^{8.5}}\\
&\quad  + \biggl[-\frac{257^1 701^1 25142051358546602833^1}{2^{17} 3^7 5^3 7^5 11^2 13^2 17^1 19^1}+\frac{2^7 107^1 151^1}{3^3 5^1 7^2}\pi ^2 + \frac{2^{10} 151^1}{3^1 5^1 7^1}\zeta (3) - \frac{2^6}{5^1}\log(2/R) \\
&\quad+\frac{53^1 467003681161^1}{3^6 5^3 7^3 11^1 13^1}\eulerlog_2(R)-\frac{2^9107^1151^1}{3^25^27^2} \eulerlog_2^2(R)\biggr]\frac{1}{R^9}+
\biggl[-\frac{ 2^3 34871408153^1}{3^5 5^3 7^3 11^1}-\frac{ 2^{11} 107^1}{ 3^2 5^2 7^1}\pi^2 \\
&\quad + \frac{2^{11}107^2}{3^25^37^2} \eulerlog_2(R)\biggr]\frac{\pi}{R^{9.5}}
+ \biggl[-\frac{21937^129041277^1138624434738172773917^1}{2^{18}3^75^57^611^313^217^219^123^1}+ \frac{2^5107^16899^1}{3^55^17^3}\pi^2 + \frac{2^86899^1}{3^35^17^2}\zeta(3) \\
&\quad - \frac{2^53^2}{5^1}\log(2/R) + \frac{157^137456616009^1}{2^1 3^6 5^2 7^4 13^1}\eulerlog_2(R)-\frac{2^7107^16899^1}{3^45^27^3} \eulerlog_2^2(R)\biggr]\frac{1}{R^{10}}+\biggl[\frac{239^12819^14306391899^1}{3^75^47^411^113^1}\\
&\quad + \frac{2^{10}107^1151^1}{3^35^27^2}\pi^2 -\frac{2^{10}107^2151^1}{3^35^37^3}\eulerlog_2(R)\biggr]\frac{\pi}{R^{10.5}} + \biggl[\frac{283^1  428059326177089343878572901960570207^1}{2^{20}  3^8  5^6  7^7  11^3  13^3  17^2  19^2  23^1}\\
&\quad-\frac{ 2^4  34871408153^1}{3^6  5^2  7^3  11^1}\pi ^2-\frac{2^{12}  107^1}{3^2  5^3  7^1}\pi ^4 -\frac{ 2^7  29^1  58564963^1}{3^4  5^3  7^2  11^1}\zeta (3) +\frac{2^{15} }{5^1}\zeta (5)-\frac{  2^3  4507^1}{3^1  5^1  7^1}\log(2/R) \\
&\quad + \left(-\frac{23^1  37^1  179^1  1871^1  7907^1  58007867^1}{2^3  3^6  5^5  7^4  11^2  13^1  17^1} + \frac{2^{12} 107^2}{3^35^27^2}\pi^2 + \frac{2^{15}107^1}{3^15^27^1}\zeta(3)\right)\eulerlog_2(R)+ \frac{2^634871408153^1}{3^55^37^311^1}\eulerlog_2^2(R) \\
&\quad - \frac{2^{14}107^2}{3^35^37^2}\eulerlog_2^3(R)\biggr]\frac{1}{R^{11}} + \biggl[\frac{1001321^13135169339^1}{2^13^75^47^513^1} + \frac{2^8107^16899^1}{3^55^27^3}\pi^2- \frac{2^8107^26899^1}{3^55^37^4}\eulerlog_2(R)\biggr]\frac{\pi}{R^{11.5}}\\
&\quad + \biggl[-\frac{23124661990883^1  53722231045521038313865859^1}{2^{21}  3^{11}  5^6  7^7  11^3  13^3  17^2  19^2  23^1} +\frac{ 2^1  2891598303407839^1}{3^8  5^3  7^4  11^1  13^1}\pi ^2+\frac{2^{11}  107^1  151^1}{3^3  5^3  7^2}\pi ^4\\
&\quad +\frac{ 2^4  461^1  541^1  114770629^1}{3^6  5^3  7^3  11^1  13^1}\zeta (3) -\frac{  2^{14}  151^1}{3^1  5^1  7^1}\zeta (5) -\frac{ 2^2  191^1  124343^1}{3^2  5^3  7^2}\log(2/R)+ \frac{2^9107^1}{3^15^27^1}\log^2(2/R)\\
&\quad+ \left(\frac{179^1  122938488234581263017407^1}{2^4  3^9  5^5  7^5  11^2  13^2  17^1  19^1} - \frac{2^{11}107^2151^1}{3^45^27^3}\pi^2 - \frac{2^{14}107^1151^1}{3^25^27^2}\zeta(3)\right) \eulerlog_2(R)\\
&\quad- \frac{2^3239^12819^14306391899^1}{3^75^47^411^113^1}\eulerlog_2^2(R) + \frac{2^{13}107^2151^1}{3^45^37^3}\eulerlog_2^3(R)\biggr]\frac{1}{R^{12}}+ \biggl[-\frac{79^189^1139709^12309147252033^1}{2^33^75^67^411^213^117^1}\\
&\quad - \frac{2^7170476104541^1}{3^65^47^311^1}\pi^2 + \frac{2^{15}107^1}{3^35^37^1}\pi^4 + \frac{2^{15}107^2}{3^25^37^2}\zeta(3)+ \left(\frac{2^7107^1185098037053^1}{3^65^57^411^1} + \frac{2^{15}107^2}{3^35^37^2}\pi^2\right)\eulerlog_2(R)\\
&\quad - \frac{2^{14}107^3}{3^35^47^3}\eulerlog_2^2(R) \biggr]\frac{\pi}{R^{12.5}}
+ \biggl[-\frac{30942707693267^1  3152956529329347926281546766946241^1}{2^{24}  3^{11}  5^7  7^7  11^4  13^3  17^3  19^2  23^2  29^1}\\
&\quad  + \frac{43^1  193^1  373088739161^1}{3^8  5^3  7^5  13^1}\pi ^2  +\frac{2^9  107^1  6899^1}{3^5  5^3  7^3}\pi ^4  +\frac{2^3  23^1  2459^1  524570909^1}{3^6  5^3  7^4  13^1}\zeta (3)  - \frac{ 2^{12}  6899^1}{3^3  5^1  7^2}\zeta (5) \\
&\quad-\frac{ 1493^1  185557^1}{2^1  3^1  5^3  7^2}\log(2/R)+ \frac{2^{16}}{3^1  5^2}\log(2/R)\eulerlog_2(R)+\frac{ 2^8  3^1  107^1}{5^2  7^1}\log^2(2/R)\\
&\quad+ \biggl(\frac{19^1  37^1  227^1  401^1  92033^1  7275821540983^1}{2^8  3^9  5^4  7^6  11^2  13^2  17^1}-\frac{2^9  107^2  6899^1}{3^6  5^2  7^4}\pi^2-\frac{ 2^{12}  107^1  6899^1}{3^4  5^2  7^3}\zeta(3)\biggr)\eulerlog_2(R)
\end{split}
\?
\<
\begin{split}
\phantom{\Upsilon_{22}} 
&\quad -\frac{2^2  47^2  300973^1  4819967^1}{3^7  5^4  7^5  13^1}\eulerlog_2^2(R) + \frac{2^{11}107^2 6899^1}{3^65^37^4}\eulerlog_2^3(R)\biggr]\frac{1}{R^{13}}+ \biggl[\frac{562320412209981988553368867^1}{2^43^{10}5^57^611^213^217^119^1}\\
&\quad + \frac{2^480284955734189^1}{3^85^37^311^113^1}\pi^2 - \frac{2^{14}107^1151^1}{3^45^37^2}\pi^4  - \frac{2^{14}107^2151^1}{3^35^37^3}\zeta(3) + \biggl(-\frac{2^4107^11213^119163^1135704273^1}{3^85^57^511^113^1}\\
&\quad - \frac{2^{14}107^2151^1}{3^45^37^3}\pi^2\biggr)\eulerlog_2(R) + \frac{2^{13}107^3151^1}{3^45^47^4}\eulerlog_2^2(R)\biggr]\frac{\pi}{R^{13.5}} + O\left(\frac{1}{R^{14}}\right).
\end{split}
\?
\end{widetext}
Here 
\<
\eulerlog_m(R) := \gamma + \log(2m/R^{1/2}),
\?
where $\gamma$ is the Euler-Mascheroni gamma constant, is the function associated with many higher-order tail terms in the PN expansion, first introduced in general by Damour, Iyer, and Nagar~\cite{DIN}, with a slightly different definition, since they use a different variable. Additionally, $\zeta$ denotes the Riemann zeta function.

The PN expansion of the $(2,2)$ mode for $\Delta U$ has quite a bit of structure that is readily apparent in its prime factorization, and the PN expansions of the other modes display similar structure. In particular, we can write most of the $\eulerlog^n_m(R)$, half-integer, and zeta function terms (including the even powers of $\pi$) in $\Upsilon_{\ell m}$ [the $(\ell,m)$ mode of $\Delta U/U$] to the orders currently known in the following form (``C'' is for ``complications'')
\begin{widetext}
\<
\label{eq:Upsilon_simp}
\begin{split}
\Upsilon^{C1}_{\ell m} &=  \biggl[e^{2\bar{\nu}_{\ell m}\eulerlog_m(R)}\biggl\{\frac{1}{2\bar{\nu}_{\ell m}}\frac{1}{R^2} - \frac{5}{12}\frac{\bar{\nu}_{\ell m}\pi^2}{R^2} + \frac{7}{3}\frac{\bar{\nu}_{\ell m}^2\zeta(3)}{R^2}+ (2m)^2\frac{\zeta(3)}{R^5} - \frac{m^2\bar{\nu}_{\ell m}}{15}\frac{\pi^4}{R^5} - (2m)^4\frac{\zeta(5)}{R^8} - \frac{\bar{\nu}_{\ell m}}{4m}\frac{\pi}{R^{0.5}} \\
&\quad + \frac{\bar{\nu}_{\ell m}^3}{24m}\frac{\pi^3}{R^{0.5}}  - \frac{m\bar{\nu}_{\ell m}}{3}\frac{\pi^3}{R^{3.5}} - 2m\bar{\nu}_{\ell m}^2\frac{\pi\zeta(3)}{R^{3.5}} + \frac{4m^3\bar{\nu}_{\ell m}}{45}\frac{\pi^5}{R^{6.5}}\biggr\} - \frac{1}{2\bar{\nu}_{\ell m}}\frac{1}{R^2}\biggr]\sum_{k=0}^\infty \frac{A_{\ell m}^{(k)}}{R^{k + \ell + 1 + \varepsilon_{\ell m}}}\\
&=  \Biggl[e^{2\bar{\nu}_{\ell m}\eulerlog_m(R)}\biggl\{\frac{1}{2\bar{\nu}_{\ell m}}\frac{1}{R^2} - \bar{\nu}_{\ell m}\left[\frac{5}{2}\zeta(2) - \frac{7}{3}\bar{\nu}_{\ell m}\zeta(3)\right]\frac{1}{R^2} + (2m)^2\left[\zeta(3) - \frac{3}{2}\bar{\nu}_{\ell m}\zeta(4)\right]\frac{1}{R^5}  - (2m)^4\frac{\zeta(5)}{R^8}\\
&\quad - \frac{\bar{\nu}_{\ell m}}{4m}\left[1 - \bar{\nu}_{\ell m}^2\zeta(2)\right]\frac{\pi}{R^{0.5}} - 2m\bar{\nu}_{\ell m}\left[\zeta(2) + \bar{\nu}_{\ell m}\zeta(3)\right]\frac{\pi}{R^{3.5}} + (2m)^3\bar{\nu}_{\ell m}\zeta(4)\frac{\pi}{R^{6.5}}\biggr\} - \frac{1}{2\bar{\nu}_{\ell m}}\frac{1}{R^2}\Biggr]\sum_{k=0}^\infty \frac{A_{\ell m}^{(k)}}{R^{k + \ell + 1 + \varepsilon_{\ell m}}}\\
&=: C^{[1]}_{\ell m}\sum_{k=0}^\infty \frac{A_{\ell m}^{(k)}}{R^k}.
\end{split}
\?
\end{widetext}
Here we have given two forms for $\Upsilon^{C1}_{\ell m}$ to better illustrate its structure;\footnote{Note that we only need to use $\Upsilon^{C1}_{\ell m}$ to simplify the $m\neq 0$ modes. The $m = 0$ modes are nonradiative, and thus already have purely rational simple integer-order PN series, with no simplification necessary. Therefore, even though $\bar{\nu}_{\ell m} = 0$ for $m = 0$, one does not need to be concerned about potential division by zero [or logarithms of zero in $\eulerlog_m(R)$] in Eq.~\eqref{eq:Upsilon_simp}.} recall that $\zeta(2) = \pi^2/6$ and $\zeta(4) = \pi^4/90$. Additionally,
\<
\bar{\nu}_{\ell m} := \nu - \ell = \sum_{k = 1}^\infty\od{\nu_\ell}{k}\frac{(2m)^{2k}}{R^{3k}},
\?
where $\nu$ is the renormalized angular momentum introduced in the MST formalism~\cite{MST, ST}. (Here we denote its dependence on $\ell$ and $m$ explicitly, which is usually not done in the literature, though we
suppress its dependence on $R$, even though we displayed the analogous dependence on $v$ in~\cite{NKJ-M_tamingPN}.) See the Appendix of Bini and Damour~\cite{BD2} for explicit expressions for $\od{\nu_\ell}{k}$, $k\in\{1,2,3\}$, where these are referred to as $\nu_{2k}(\ell)$. Note also that $\od{\nu_2}{1} = -107^1/2^13^15^17^1$, which explains the appearance of factors of $107$ in many places in the prime factorization of $\Upsilon_{22}$
in Eq.~\eqref{eq:Upsilon22}. In fact, $\nu$ (along with its analogue for $\ell \to -\ell - 1$) gives many of the leading logarithms in the homogeneous solutions of the Regge-Wheeler equation, as noted in Sec.~II~B of~\cite{KOW} (some similar results for the Teukolsky equation are also implicit in the results of~\cite{NKJ-M_tamingPN}). Additionally, $-\ri\nu$ is the monodromy of the radial Teukolsky equation about the irregular singular point at infinity, as is mentioned in~\cite{CLMR1}. One also
sees $\od{\nu_\ell}{1}$, multiplied by a rational with small prime factors, appearing in the coefficients of integrals involving an $\ell$ multipole in the standard PN calculation of the next-to-leading two half-integer terms in $\Delta U$ in~\cite{BFW2}; cf.\ their Eqs.~(3.14)--(3.18) and (4.7)--(4.11) with the values for $\od{\nu_\ell}{1}$, $\ell\in\{2,3,4\}$ given in Table~I in~\cite{NKJ-M_tamingPN}. Finally, the general form of $\od{\nu_\ell}{1}$ appears in
the coefficient of $\log\tau_0$ (where $\tau_0$ is the constant associated with the regularization parameter $r_0$, not to be confused with the $r_0$ in this paper) in post-Newtonian expressions for all the mass-type radiative multipole moments; cf.\ Eq.~(3.9) in~\cite{FBI} and Eq.~(A2) in~\cite{BD2}. This coefficient was derived by Blanchet and Damour in the Appendix of~\cite{BD_tails} using methods that differ from both the continued fraction method of MST~\cite{MST} and the monodromy method of Castro~\emph{et al.}~\cite{CLMR1}.
We also define
\<\label{eq:vareps}
\varepsilon_{\ell m} :=
\begin{cases}
0 &\text{if $\ell + m$ is even},\\
1 &\text{if $\ell + m$ is odd}.
\end{cases}
\?
The $A_{\ell m}^{(k)}$ coefficients are rational and are given by the coefficients of the $\eulerlog_m(R)$ terms. While we might expect there to be contributions to the eulerlog terms that are not part of this simplification
starting at $9$PN, by analogy with the remainder of the $S_{\ell m}$ factorization of the modes of the energy flux from~\cite{NKJ-M_tamingPN}, it appears that this is not the case, since we see the same structure in the remainder with this choice for the $A_{\ell m}^{(k)}$ coefficients as for the  $S_{\ell m}$ factorization of the modes of the energy flux to all the orders we have considered.

If we apply this simplification to the $(2,2)$ mode, then we have %\nkjm{Perhaps don't give the prime factorizations for $A_{22}^{(k)}$...}
\begin{widetext}
\<
\label{eq:A22}
\begin{split}
\sum_{k=0}^\infty \frac{A_{22}^{(k)}}{R^k} &= -\frac{2^7}{5^1} + \frac{2^6151^1}{3^15^17^1}\frac{1}{R} + \frac{2^46899^1}{3^35^17^2}\frac{1}{R^2} - \frac{2^3 1529121911^1}{3^45^37^211^1}\frac{1}{R^3} + \frac{53^1 467003681161^1}{3^65^37^311^113^1}\frac{1}{R^4} + \frac{157^1 37456616009^1}{2^13^65^27^413^1}\frac{1}{R^5}\\
&\quad -\frac{23^137^1179^11871^17907^158007867^1}{2^33^65^57^411^213^117^1}\frac{1}{R^6} + \frac{179^1  122938488234581263017407^1}{2^4  3^9  5^5  7^5  11^2  13^2  17^1  19^1}\frac{1}{R^7}\\
&\quad + \frac{19^1 37^1 227^1 401^1 92033^1   7275821540983^1}{2^8 3^9 5^4 7^6 11^2 13^2 17^1}\frac{1}{R^8} + \cdots
\end{split}
\?
which yields
\<\label{eq:U22_simp} 
\begin{split}
\Upsilon_{22} - \Upsilon^{C1}_{22} &=\sum_{k = 1}^{13}\frac{\alpha_{22}^{(k)}}{R^k} + \left[-\frac{2^6}{5^1}\frac{1}{R^9}  -\frac{2^53^2}{5^1}\frac{1}{R^{10}} - \frac{2^34507^1}{3^15^17^1}\frac{1}{R^{11}} - \frac{2^2191^1124343^1}{3^25^37^2}\frac{1}{R^{12}} - \frac{1493^1  185557^1}{2^1  3^1  5^3  7^2}\frac{1}{R^{13}}\right]\log(2/R)  \\
&\quad + \frac{2^{16}}{3^1  5^2}\frac{\log(2/R)\eulerlog_2(R)}{R^{13}} + \left[\frac{2^9107^1}{3^15^27^1}\frac{1}{R^{12}} + \frac{2^8  3^1  107^1}{5^2  7^1}\frac{1}{R^{13}}\right]\log^2(2/R) + \biggl[-\frac{2^{10}107^1}{3^25^27^1}\frac{1}{R^{12}} \\
&\quad-\frac{ 2^9  11^1 13^2}{3^2  5^2  7^1}\frac{1}{R^{13}}\biggr]\pi^2 + \left[-\frac{2^{11}}{5^1}\frac{1}{R^{12}}-\frac{2^{10}  3^2}{5^1}\frac{1}{R^{13}}\right]\zeta(3)  - \frac{2^{13}}{3^15^2}\frac{\pi}{R^{11.5}} - \frac{2^{12}19^1}{3^25^27^1}\frac{\pi}{R^{12.5}} -\frac{2^{10} 11^1 673^1}{3^4 5^2 7^1}\frac{\pi}{R^{13.5}}\\
&\quad 
+ \left\{\frac{1}{R^{14}}\text{ and higher terms that we do not yet know all of}\right\} + \biggl[-\frac{2^9 240013637^1}{3^3 5^4 7^2 11^1}+ \frac{2^{17}  107^1}{3^2  5^3  7^1}\eulerlog_2(R)\\
&\quad +\frac{2^{16}  107^1}{3^1  5^3  7^1}\log(2/R) - \frac{2^{17}}{3^2  5^2}\pi^2\biggr]\frac{\pi}{R^{14.5}}- \frac{2^{13}107^2}{3^35^37^2}\frac{\log^3(2/R)}{R^{15}} +  O\left(\frac{1}{R^{15.5}}\right),
\end{split}
\?
\end{widetext}
where $\alpha_{22}^{(k)}\in\Q$. Here, to simplify the remainder, we have not used $\od{\nu}{3}$ in $\Upsilon^{C1}_{22}$, but rather the expression for the $1/R^9$ piece of the $1/R$ expansion of $\nu$ that is valid for all $\ell > 2$, without the additional
piece that only contributes for $\ell = 2$ (for positive $\ell$). Specifically, this is the expression for $\nu_6(\ell)$ given in the Appendix of Bini and Damour~\cite{BD2} with the final $c_6$ term omitted. We omit this term
because $c_6 = 2^17^1/3^15^1107^1$, and we do not actually see such factors of $107$ (or any other anomalously large primes) in the denominators of the PN expansion of fluxes or gauge-invariant self-force quantities
available to date. % \nkjm{Give caveats that we haven't looked at Fujita's new results since they aren't yet available electronically---or just ask Ryuichi about this... Also, we may want to be careful saying ``observable quantity''---
%perhaps ``gauge-invariant quantity'' would be better... \nkjm{Is it clear that $\nu$ is gauge-dependent?}
For instance, if we had used $\od{\nu}{3}$ instead of the general $\ell > 2$ expression in $\Upsilon^{C1}_{22}$, then
the $11.5$PN and higher half-integer coefficients in the remainder would have had more complicated expressions (with factors of $107$ in the denominator). In particular, the $11.5$PN coefficient in the remainder would be $2^{12}6737^1\pi/3^25^27^1107^1$. However, we do see such factors of $107$, and other large primes from the numerator of the $\od{\nu}{1}$ in the denominators of certain terms in the factorizations of the modes of the energy flux at infinity given in~\cite{NKJ-M_tamingPN}, which is not surprising, since the full $\nu$ is used in these simplifications.

Indeed, if one looks at the coefficients of $\eulerlog_m(R)$ in the PN expansion of the logarithms of the modes of the
energy flux at infinity (for a point particle in a circular orbit around a Schwarzschild black hole), then these coefficients give the coefficients of the $1/R$ expansion of $\nu$ up to the point at which the first departure from the general $\ell$-behavior occurs (at $1/R^{3(\ell + 1)}$): Specifically, if $\eta_{\ell m}$ denotes the $(\ell, m)$ mode of this energy flux, the coefficient of $\eulerlog_m(v)v^{6n}$ in $\log\eta_{\ell m}$ is $2\od{\bar{\nu}_{\ell m}}{n}(2m)^{2n}$ for $n < \ell + 1$. This behavior is to be expected, given the action of the $S_{\ell m}$ factorization from~\cite{NKJ-M_tamingPN}; $n = \ell + 1$ is the point at which the $-\nu - 1$ portion of $B^\text{inc}_{\ell m\omega}$ in the MST formalism starts to contribute $\eulerlog_m(v)$ terms to $\eta_{\ell m}$ [cf.\ Eq.~(19b) in~\cite{NKJ-M_tamingPN} and the discussion in Sec.~IV of that paper]. What is striking is that the coefficient of $\eulerlog_m(v)v^{6(\ell + 1)}$ is given by the general-$\ell$ expression for the PN coefficients of $\nu$. For an illustration of all of this for the $(2,2)$ mode, compare the expression for $\log\eta_{\ell m}$ in Eq.~(32) in~\cite{NKJ-M_tamingPN} with the expressions for the PN expansion of $\nu$ given in the Appendix of Bini and Damour~\cite{BD2}. However, for $n = \ell + 2$ things are more complicated, since the expression for the PN coefficient for a general $\ell$ develops a pole, whose residue seems to have nothing to do with the difference between the coefficients of $\eulerlog_m(v)v^{6n}$ in $\log\eta_{\ell m}$ and the PN coefficients of $\nu$. In particular, this residue is quite simple and has no large prime factors.

The simplification we introduce here is in many ways analogous to the $S_{\ell m}$ factorization of the modes of the energy flux at infinity introduced in~\cite{NKJ-M_tamingPN}, and likely has a similar expression in terms of
gamma functions, where the current expression is just low-order terms in its PN expansion. However, while it is reasonably easy to read off the $S_{\ell m}$ factorization from the MST expression for the modes of the energy flux, it is far less easy to ascertain the similar full expression for this simplification of the modes of $\Delta U$, except for the $e^{2\bar{\nu}_{\ell m}\eulerlog_m(R)}$ piece, as discussed in the Appendix. Note also that the $S_{\ell m}$ factorization is applied to the entire mode (by division), while here we only subtract off a portion of the expansion with the simplification.

There is also some notable structure in the remainder (in particular all the factors of $107$ in the prime factorization), and it appears that the powers of $\log(2/R)$ and the $\zeta(3)$ terms in the remainder can all be derived
from a single series, akin to $\Upsilon^{C1}_{\ell m}$. By analogy with $\Upsilon^{C1}_{22}$ and the $V_{\ell m}$ factorization of the modes of the energy flux from~\cite{NKJ-M_tamingPN} (though the $V_{\ell m}$ factorization does not remove some of the terms that are analogous to those considered here, so the analogy is far from exact), we conjecture that it has the form
\<\label{eq:Upsilon_simpC2}
\begin{split}
\Upsilon_{\ell m}^{C2} &= \left\{e^{2\bar{\nu}_{\ell m}\log(2/R)}\left[\frac{1}{2\bar{\nu}_{\ell m}} + \frac{2^3m^2\zeta(3)}{R^3}\right] - \frac{1}{2\bar{\nu}_{\ell m}}\right\}\\
&\quad \times\sum_{k=0}^\infty\frac{B_{\ell m}^{(k)}}{R^{k+5+2\ell+\epsilon_{\ell m}}}\\
&=: C^{[2]}_{\ell m}\sum_{k=0}^\infty\frac{B_{\ell m}^{(k)}}{R^k},
\end{split}
\?
where
\<
\label{eq:B22}
\begin{split}
\sum_{k=0}^\infty\frac{B_{22}^{(k)}}{R^k} &= -\frac{2^6}{5^1} - \frac{2^53^2}{5^1}\frac{1}{R} - \frac{2^34507^1}{3^15^17^1}\frac{1}{R^2}\\
&\quad - \frac{2^2191^1124343^1}{3^25^37^2}\frac{1}{R^3} - \frac{1493^1185557^1}{2^13^15^37^2}\frac{1}{R^4}+ \cdots.
\end{split}
\?
and we give the expressions for the other modes to the order we know them in the electronic Supplemental Material~\cite{J-MSW_suppl}.
However, note that we do not yet know the expansion of the individual modes to high enough orders to be able to check whether many of the predicted terms appear, and whether the coefficient of $\log(2/R)/R^{14}$ also gives the coefficients of the other higher-order terms this expression suggests it will. Nevertheless, we are able to check some of these predictions for the first appearance of a given power of a logarithm in the $(2,2)$ mode using our
results for the higher-order PN coefficients of the full $\Delta U$ and the rest of the simplification, as discussed in Sec.~\ref{sec:fit}. Additionally, we obtain a few less direct checks on more of these predictions for the $(2,2)$ mode and others from the simplifications of the remainders of other logarithmic terms in the full $\Delta U$. Moreover, there is a very similar structure in the remainder of the $S_{\ell m}$ factorization of the modes of the energy flux at infinity~\cite{NKJ-M_tamingPN}, lending further support to this
conjectured form. In general, these similarities between the structures that can be simplified for the energy flux and $\Delta U$ are not surprising, since they likely all come from tail effects.

%--------------------------
\subsection{Applying PSLQ to the coefficients of the PN expansion of the modes of $\Delta U$}
%--------------------------

We now outline the general method we use to obtain the analytic forms of the coefficients of the PN expansion of the modes of $\Delta U$. First, we note that the form of the PN expansion of the modes of the energy flux generally provides a good guide to the growth of complexity of the terms in the expansion of the modes of $\Delta U$, in particular concerning the appearance of terms that we are not able to remove using the simplification---cf.\ the discussion in~\cite{NKJ-M_tamingPN}. Next, we note that the individual retarded $(\ell, m)$ modes of $\Delta U$ have the following general structure
\begin{widetext}
%\begin{widetext}
\<\label{eq:dU_form}
\begin{split}
&\quad \sum_{n = 1+ \varepsilon_{\ell m}}^\infty \frac{A^{(0)}_n}{R^n} + \sum_{n = 3+\ell + \varepsilon_{\ell m}}^\infty \frac{A^{(1)}_n\eulerlog_m(R)}{R^n} + \sum_{n = 6+\ell + \varepsilon_{\ell m}}^\infty \frac{A^{(2)}_n\eulerlog^2_m(R)}{R^n} +\cdots\\
&\quad+\pi\left[\sum_{n = 4+\ell + \varepsilon_{\ell m}}^\infty\frac{B^{(0)}_n}{R^{n+1/2}}+\sum_{n = 7+\ell + \varepsilon_{\ell m}}^\infty\frac{B^{(1)}_n\eulerlog_m(R)}{R^{n+1/2}}+\sum_{n = 10+\ell + \varepsilon_{\ell m}}^\infty\frac{B^{(2)}_n\eulerlog^2_m(R)}{R^{n+1/2}} + \cdots\right]\\
&\quad + \sum_{n = 5+2\ell+\varepsilon_{\ell m}}^\infty \frac{C^{(0)}_n\log(2/R)}{R^n} + \sum_{n = 8+2\ell+\varepsilon_{\ell m}}^\infty \frac{C^{(1)}_n\log^2(2/R)}{R^n} + \cdots\\
&= \sum_{n = 1+ \varepsilon_{\ell m}}^\infty \frac{A^{(0)}_n}{R^n} + \sum_{k=0}^\infty\biggl[\sum_{n = 3k+3+\ell + \varepsilon_{\ell m}}^\infty \frac{A^{(k+1)}_n\eulerlog^{k+1}_m(R)}{R^n}+\pi\sum_{n = 3k+4+\ell + \varepsilon_{\ell m}}^\infty\frac{B^{(k)}_n\eulerlog^k_m(R)}{R^{n+1/2}}\\
&\quad +
\sum_{n = 3k+5+2\ell+\varepsilon_{\ell m}}^\infty \frac{C^{(k)}_n\log^{k+1}(2/R)}{R^n}
+ \cdots\biggr].
\end{split}
\?
\end{widetext}
[Recall that $\varepsilon_{\ell m}$ is defined in Eq.~\eqref{eq:vareps}.]
In particular, note that the individual modes are purely rational integer-order PN series until the first appearance of the logarithm, where they start to have transcendental contributions, as well. The transcendentals and the logarithm first only appear together in the form of $\eulerlog_m(R)$ in the integer-order terms, but then, starting with the $1/R^{6+\ell +\epsilon_{\ell m}}$ term (i.e., the same order at which $\eulerlog^2$ terms start to appear), they also get $\pi^2$ and $\zeta(3)$ terms, where $\zeta$ is the Riemann zeta function. Much of the increase of complexity is described by the simplifications [Eqs.~\eqref{eq:Upsilon_simp} and~\eqref{eq:Upsilon_simpC2}], though there are a few logarithms, transcendentals,\footnote{Note that $\gamma$ and $\zeta$ evaluated at odd integers are not known to be transcendental, or in most cases even irrational. However, they are all strongly conjectured to be transcendental, so we shall refer to them as such.} and half-integer terms that the simplifications do not remove, just as found for the energy flux in~\cite{NKJ-M_tamingPN}, which should be expected, from the form of the expressions used to obtain both quantities. In particular, the expression in Eq.~\eqref{eq:dU_form} does not include the appearance of the $\eulerlog_2(R)\log(2/R)/R^{13}$ and $\pi\log(2/R)/R^{14.5}$ terms that are known in the $(2,2)$ mode of $\Delta U$ (which are also not given by either of the simplifications).

We apply the PSLQ integer relation algorithm~\cite{FB,FBA} in its implementation as the FindIntegerNullVector function in {\sc{Mathematica}} to obtain analytic forms for the PN coefficients of the retarded $(\ell, m)$ modes of $\Delta U$.\footnote{Note that we shall often use the name PSLQ as a shorthand for the FindIntegerNullVector function.} Specifically, if one inputs a vector of decimals to the PSLQ algorithm, it returns the (nonzero) vector that is orthogonal to the input vector and has a small ($L^2$) norm. One can thus apply PSLQ to identify the analytic form of a number from a sufficiently accurate decimal representation, if one knows (or has an educated guess for) the transcendental numbers [here, for instance, $\pi$, $\log(2)$, $\gamma$, $\zeta(3)$, etc.]\ present in the analytic form. This is particularly simple when the number one will obtain is a linear combination of the transcendentals with rational coefficients, as is the case here, where the vector in question is simply the decimal expansion of the number to be identified, along with 1, and any transcendentals thought to be present. Of course, PSLQ will give an output for any vector, but the outputs that do not correspond to a true relation are almost always large and ``ugly-looking'' for a sufficient number of digits, while the true vector will have a certain ``nice-looking'' structure (which we will discuss further later). %\nkjm{Put some of this in the introduction?}

%...........................
\subsection{An example: Obtaining the analytic form of the $\beta_7$ coefficient of the full $\Delta U$ from the decimal form given in SFW}
%...........................

As an example, we consider obtaining the $\beta_7$ coefficient [i.e., the coefficient of the $\log(R)/R^8$ term in $\Delta U$] from the numerical value given in SFW, as was reported there (and confirmed by the analytic calculation in~\cite{BD3}). This is the coefficient of $\log(R)$ at the first order where there is a $\log^2(R)$ term, so, by analogy with the transcendentals appearing in the nonlogarithmic term at the first appearance of $\log(R)$, we expect to have $\gamma$ and $\log(2)$ terms here. As we saw in the expression of the structure of the PN expansion of the modes in terms of eulerlogs above, this linking of $\log(R)$, $\log(2)$, and $\gamma$ is a generic feature, though it is broken at high orders by the appearance of the $\log(2/R)$ terms. Indeed, since this is the first appearance of $\log^2(R)$, and thus only comes from the $(2,2)$ mode, we can actually subtract off the $\log(2)$ and $\gamma$ terms and only have to obtain the rational piece using PSLQ. However, we shall first consider the case of using PSLQ to obtain the full term, since this is how we initially obtained it.
%Ordinarily one needs to look at the individual modes separately to be able to do this, which is why we usually consider them separately.

\begin{figure}[htb]
\begin{center}
\epsfig{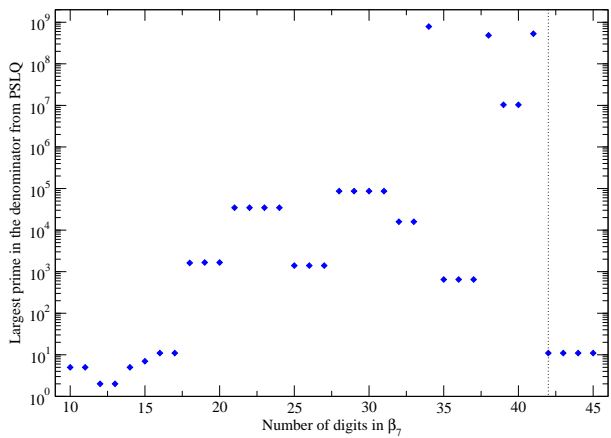}
\end{center}
\caption{\label{fig:beta7} The largest prime in the denominator of the expression for $\beta_7$ returned by {\sc Mathematica}'s FindIntegerNullVector function applied to the vector $\{\beta_7, 1, \gamma, \log(2)\}$
for $\beta_7$ evaluated to varying numbers of digits from $10$ to $45$. The vertical dotted line marks the point at which this function returns an accurate analytic expression for $\beta_7$.}
\end{figure}

Starting from
%\begin{widetext}
\<
\begin{split}
\beta_7 &=  536.4052124710242868717895394750389112702062\\
&\quad\;69552321207927883360240368736326766131833\ldots,
\end{split}
\?
%\end{widetext}
which is taken directly from Table~I in SFW, we can apply PSLQ in the form of {\sc{Mathematica}}'s FindIntegerNullVector function to the vector $\{\beta_7,1,\gamma,\log(2)\}$ and obtain the expression of
\<
\beta_7 = \frac{5163722519}{5457375}-\frac{109568}{525} \gamma -\frac{219136 }{525}\log (2)
\?
with at least $42$ digits. (\emph{Nota bene}: We find that the final digit given by SFW is incorrect and should be a $6$.) We were able to reject the expressions produced by smaller numbers of digits since they lead to anomalously large prime factors in the denominator (i.e., the term in the vector returned by PSLQ corresponding to $\beta_7$ itself), except if one only evaluates $\beta_7$ to a very small number of digits, of course: See Fig.~\ref{fig:beta7} (cf.\ Fig.~5 in~\cite{BB_Mathematics}, which shows an alternative method for detecting a likely true relation using PSLQ by looking at the size of the smallest entry in the vector versus the number of iterations of the algorithm, which is not information available when using {\sc{Mathematica}}'s FindIntegerNullVector). We shall discuss this method of looking at the prime factorization further in Sec.~\ref{sec:checks}.

Interestingly enough, the minimum number of digits required to obtain this expression accurately with FindIntegerNullVector is somewhat dependent on the
order of the terms in the vector. For instance, if we consider instead the vector $\{\beta_7,\log(2),1,\gamma\}$, we only need $39$ digits to obtain an accurate analytic form. If we scale by the denominator of $\beta_6$ (which is
$575$), i.e., consider the vector $\{575\beta_7,1,\gamma,\log(2)\}$ we also only need $39$ digits; this sort of scaling is much more effective at higher orders where the denominators are much larger, and can decrease the
minimum number of digits required to obtain an accurate expression by $27$ digits or more. If we subtract off the $\gamma$ and $\log(2)$ terms using the expectations from the $\eulerlog_2^2(R)$ form of the $\log^2(R)$
coefficient at this order, then we only need $22$ digits to obtain the resulting $\bar{\beta}_7$ (here we use the vector $\{\bar{\beta}_7,1\}$, of course): While FindIntegerNullVector returns the correct result with between $12$ and $15$ digits in this case, it then returns an erroneous relation when one uses between $16$ and $21$ digits before returning to the correct result for $22$ digits and above. We have not observed such unusual behavior in our other determinations. Here scaling by $575$ actually \emph{increases} the minimum number of digits required for an accurate result by $1$, which can sometimes be the case when one is scaling by a relatively small number, as here.

%..............................
\subsection{Combining together the values at different radii to increase the number of digits known}
%..............................

For the low-order coefficients, which are purely rational integer-order PN coefficients, we can apply PSLQ directly to an appropriate number of digits at a given radius (e.g., for $R = 10^{50}$, one can expect to get at least $\sim 40$ accurate digits at a given order for the leading term). If one can identify the rational represented using PSLQ with this number of digits, then one can subtract it off and proceed to the next order. For the higher-$\ell$ modes, the purely rational coefficients persist to high enough orders and are large enough that one needs more than the number of digits provided by merely evaluating $\Delta U$ at $R = 10^{70}$, the largest radius we consider. In such cases, and also when we need to consider cases with logarithms and transcendentals at higher orders, we combine together the values from several radii (up to as many as $15$ radii for certain high-order pieces). The expressions we use for this purpose are long and unilluminating, but we give a simple example here to illustrate the method.

Let us assume that we are at a point in the computation where we expect that the first few terms of the PN expansion of the mode we are considering to look like
 \<
 \begin{split}
S_N(R) &= \frac{\alpha_{N,0}}{R^N} + \frac{\alpha_{N+1,0} + \alpha_{N+1,1}\log(R)}{R^{N+1}} + \frac{\alpha_{N+2,0}}{R^{N+2}}\\
&\quad + O(R^{-N-3})
 \end{split}
 \?
 (taking the $R^{-N-2}$ term to have no logarithms, since we are just interested in its overall scaling, even though this will never be the case in this sort of situation in actuality),
 and we wish to obtain $\alpha_{N,0}$ to $\sim 2k$ digits. Now, since we know the value of $S_N(R)$ at $R=10^k$ for a range of integers $k$, we can combine together the values of $S_N(R)$ at the radii $R = 10^k$, $10^{k+p}$, and $10^{k+q}$ (with $p,q\in\N$ such that $10^{k+p}$ and $10^{k+q}$ give radii at which we know the value of $S_N$), giving
 \begin{widetext}
 \<\label{eq:SN_example}
 \alpha_{N,0} = 10^{kN}\frac{(q-p)S_N(10^k) - q 10^{(N+1)p}S_N(10^{k+p}) + p 10^{(N+1)q}S_N(10^{k+q})}{q - p -q10^p + p10^q} - \underbrace{\frac{q - p - q10^{-p} + p10^{-q}}{q - p -q10^p + p10^q}\frac{\alpha_{N+2,0}}{10^{2k}}}_\cR.
 \?
 \end{widetext}
Here the remainder $\cR$ gives a small enough correction that the first term will give $\alpha_{N,0}$ to $\sim 2k$ digits, provided that $\alpha_{N+2,0}$ is not much larger than $\alpha_{N,0}$, which will usually be the case. (Note that we have neglected the rest of the remainder, whose leading term goes as $10^{-3k}$. We also have not included half-integer terms in the remainders for simplicity, though the presence of a half-integer term before the given remainder term will, of course, reduce the number of digits one obtains from the expression.)

We obtained the particular linear combination given in Eq.~\eqref{eq:SN_example} by considering $S_N(10^k) + AS_N(10^{k+p}) + BS_N(10^{k+q})$ and fixing the coefficients $A$ and $B$ by demanding that the resulting expression does not contain the two $R^{-N-1}$ terms (viz., $\alpha_{N+1,0}$ and $\alpha_{N+1,1}$). One then solves the resulting expression for $\alpha_{N,0}$ to obtain Eq.~\eqref{eq:SN_example}.
One derives the more involved expressions for more complicated cases with more terms and more radii in the same way by solving a linear system, which {\sc{Mathematica}} will do quite efficiently. For the determinations
reported in this work, we needed at most $1240$ digits and $15$ radii, which we used to determine the coefficient of the $\log^2(R)/R^{17.5}$ term in the $(2,2)$ mode [and implicitly check the prediction of the simplification for the coefficient of the $\log^3(R)/R^{17.5}$ term]; removing these terms was necessary to obtain the nonlogarithmic part of $12$PN coefficient of the $(2,2)$ mode.

%.................
\subsection{Overview of our method for obtaining analytic forms of the coefficients of the modes of $\Delta U$}
%.................

Our general approach for determining analytic forms of the PN coefficients of the modes of $\Delta U$ is first to obtain the coefficient of the highest power of $\log(R)$ present at a given order, which will always be rational (or a rational times $\pi$, for the half-integer terms), and will always come from the corresponding power of $\eulerlog_m(R)$, where $m$ is the mode's degree (i.e., its magnetic quantum number). Indeed, once we have obtained the coefficients of the first three $\log(R)$ terms in the PN expansion of a given mode, we are able to predict the coefficients of all of the highest powers of $\log(R)$ (and, in fact, much more) using the simplification. Thus, while an individual PN coefficient of the $(2,2)$ mode of $\Delta U$ can have as many as $17$ transcendentals at the relatively high PN orders we are considering, we have to use at most $5$ transcendentals in the vector to which we apply PSLQ (for the $10$PN nonlogarithmic term), since the coefficients of the remaining transcendentals are predicted by
the simplification, or given by the coefficients of higher powers of logarithms at that order. (We only need $4$ transcendentals for the nonlogarithmic piece at $12$PN, despite its more complicated structure, since at
this point in the calculation we have removed some of the transcendentals that entered at $10$PN using the simplification.)

Once we have obtained (or---more often---checked the simplification's prediction for) the coefficient of the highest power of $\log(R)$ at a given PN order, we then subtract off the appropriate rational times a power of $\eulerlog_m(R)$ and proceed to the lower powers of $\log(R)$, which have a more complicated
structure. At the orders where there are powers of $\log(2/R)$ present, we still subtract off the putative contribution as if the $\log^n(R)$ term came solely from a $\eulerlog_m^n(R)$ term, and then include the appropriate piece
when obtaining the coefficient of the next lower power of $\log(R)$ to account for the presence of the $\log^n(R)$ term. For instance, when the $\log(R)$ term comes from $a\eulerlog_m(R) + b\log(2/R)$, so we subtract off
$(a + 2b)\eulerlog_m(R)$, taking the $\log(R)$ term to come solely from an $\eulerlog_m(R)$ term, we thus include the remaining transcendental, viz., $2\gamma + \log(2) + 2\log(m)$, in the vector to which we apply PSLQ when obtaining the coefficient of the nonlogarithmic term at this PN order.

The only exception to this procedure occurs when we can predict the coefficient of the $\eulerlog_m^n(R)$ contribution from the simplification (necessarily for $n\geq 2$, since the coefficients of the $\eulerlog_m(R)$ terms are inputs to the simplification, and thus not predicted by it), in which case we simply obtain the coefficient of $\log^n(2/R)$ directly from the coefficient of $\log^n(R)$. Additionally, at $12$PN in the $(2,2)$ mode, things are quite complicated, since we have to disentangle contributions from $\eulerlog_2(R)$, $\eulerlog_2^2(R)$, $\eulerlog_2(R)\log(2/R)$, $\log(2/R)$, and $\log^2(2/R)$ terms. Here we just
subtract off the $\log^2(R)$ coefficient we obtained as a $\log^2(R)$ term and then include $\gamma + \log(2)$ and $\log(2)$ in the vector to which we apply PSLQ to obtain the $\log(R)$ coefficient. The coefficients of $\gamma + \log(2)$ and $\log(2)$ in the $\log(R)$ coefficient then let us predict the coefficients of certain $\gamma\log(2)$ and $\log^2(2)$ contributions in the nonlogarithmic coefficient, so we need to include only $2\gamma + 3\log(2)$ and $\gamma^2 + 3\gamma\log(2) + (9/4)\log^2(2)$ in the vector to which we apply PSLQ.
%Once we had checked enough of the predictions of the simplification to be confident that it, we could even check the prediction of the simplification implicitly by making sure that we were able to obtain an appropriately simple coefficient for the next lower power of $\log(R)$.

For the more complicated terms, we use the modes of the energy flux at infinity as a guide to the transcendentals we expect to be present that are not already predicted by the form of the simplification we have determined at a
given order. (These modes have been calculated to $22$PN by Fujita~\cite{Fujita22PN}, with simplified and factorized forms given in~\cite{NKJ-M_tamingPN}.) Once we have obtained a new transcendental  (or other new contribution that looks as if it is part of the simplification) at a given order\footnote{Here we consider not the absolute PN order, but the relative PN order past the first appearance of an $\eulerlog$ term, since a term at such a relative PN order will have the same complexity in all modes, while the complexity at a given absolute PN order decreases with increasing $\ell + \varepsilon_{\ell m}$.} in a few modes, we conjecture its general appearance in the simplification and then check it using a few other modes. Again, comparison with the analogous $S_{\ell m}$ factorization of the modes of the energy flux from~\cite{NKJ-M_tamingPN} is useful in determining what likely comes from the simplification, though the structures are not exactly the same. We also may need to obtain simpler higher-order terms (i.e., higher powers of logarithms at higher PN orders, or half-integer terms, which are simpler than integer-order terms at comparable orders) first, in order to remove sufficient terms from the series so that we can obtain the expression to enough digits with a relatively small number of radii. 

Finally, as mentioned above, we generally scale by the denominator of the rational term at the previous PN order (i.e., the current PN order minus $1$), since we find that this significantly reduces the number of digits
required to make an accurate determination of the analytic form. One exception to this is any case where the simplification predicts most of the term [e.g., up to a $\log^n(2/R)$ contribution, or the additional contributions to the nonlogarithmic part of half-integer terms at higher orders]. In this case, we will not scale at all, or scale by 
the denominator of the similar addition at the previous PN order, as these additional terms have significantly simpler denominators than the full coefficient. With these scalings, we need at most $249$ digits to successfully obtain an analytic form. This maximum is needed for the nonlogarithmic $1/R^{13}$ term of the $(2,2)$ mode, the most complicated coefficient of a mode we consider.

For the modes with $\ell\geq4$, we also scale by a high power of $m$ (as high as $12$ for $\ell = 9$ and $10$) when determining the (linear) $\log(R)$ and half-integer terms, since such high powers of $m$ occur there. For these modes, we are able to use the simplification to make the determination of the analytic form of the PN coefficients mostly automatic (in the sense that we only have to choose the number of digits used and various scalings, and verify that the results satisfy all the consistency checks we discuss below). Note also that we are only considering the linear $\log(R)$ terms here, since all the higher powers of $\log(R)$ in these modes are given by the simplification, to the order we are currently working.

%.........................
\subsection{Checks on the output of PSLQ}
\label{sec:checks}
%.........................

When we are performing these PSLQ determinations, it is important to ensure that one has sufficient accuracy (and the correct transcendentals in the vector) so that the form returned by PSLQ is reliable. Besides the basic test of making sure that the analytic form does not change as one increases the number of digits, up to the maximum number that are expected to be given accurately by the combination of radii being used, a very stringent test is generally making sure that the denominators do not have any large prime factors. Such \emph{smooth numbers} are distributed relatively sparsely among large numbers when the largest prime allowed is relatively small (e.g., smaller than the logarithm of the large number): See, e.g., Granville's review~\cite{Granville} for a discussion of the
properties of smooth numbers.

In particular, Granville gives an upper bound on the smooth number counting function %$\Psi(x,y)$
in his Eq.~(1.23) that gives an easy way to see how unlikely it is for a randomly selected $d$ digit number to have all its prime factors less than $p$. This probability will be less than
\<
\frac{1}{10^{d+1}-10^d}\binom{\lfloor (d+1)\log(10)/\log(2)\rfloor + \pi(p)}{\pi(p)},
\?
%\nkjm{Write the factor in the denominator as $9\times10^d$?}
where $\binom{\cdot}{\cdot}$ is the binomial coefficient, $\lfloor\cdot\rfloor$ is the floor function, and $\pi(p)$ is the number of primes $\leq p$. This probability is generally \emph{extremely} small for the cases in question. For instance, the denominators of the purely rational terms at $12$ and $12.5$PN each have $32$ digits, but their largest primes are both $19$, for which the probability is less than $10^{-21}$, for a randomly selected 32 digit number. %Computation with an asymptotic result (with poorly controlled remainders).

In addition to making sure that the denominator is a smooth number, other consistency tests include checking that the result is insensitive to small changes in the prime factorization of the overall scaling, or computing the coefficients of $U^\alpha\Delta U$ for different $\alpha$ (e.g., $\alpha = -1/2$ and $\alpha = +1/2$, though we used other values, as well) and making sure that the results are consistent. One also expects that terms that have only recently started to appear in the expansion will have simpler forms than those that have been present in the expansion for longer, which also allows one to reject some spurious expressions. In particular, one expects to see powers of the characteristic large prime from the numerator of the first PN coefficient of $\nu$ (cf.\ Table~I in~\cite{NKJ-M_tamingPN}) in such terms that have just started to appear.

Finally, if one happens to have many digits for a given term, other consistency checks include adding on other transcendentals to the vector and making sure that PSLQ gives zero coefficients for them, or obtaining the result without subtracting off some of the known results from the simplification (particularly if this is a term one already has to include in the vector anyway, since the simplification only gives a portion of it). All these checks, plus the overall consistency check that we continue to obtain expressions of the expected form for the various modes (given the simplification and the form of the modes of the energy flux) as we continue to high orders, give us high confidence that our analytic results for the modes (and, as described later, the full $\Delta U$) are indeed the true ones. This confidence was recently confirmed by the exact agreement of our results with the $20.5$PN results of Kavanagh, Ottewill, and Wardell~\cite{KOW}, obtained completely analytically.

%%%%%%%%%%%%%%%%
\section{Terms predicted by the simplifying factorization to arbitrarily high orders}
\label{sec:simp}
%%%%%%%%%%%%%%%%

The two simplifications we have found also predict certain higher-order logarithmic and half-integer terms, extending to arbitrarily high orders, since we assume that the $e^{2\bar{\nu}_{\ell m}\eulerlog_m(R)}$ and $e^{2\bar{\nu}_{\ell m}\log(2/R)}$ portions of the simplifications hold to all orders, as we expect them to, since the similar $S_{\ell m}$ and $V_{\ell m}$ factorizations found for the energy flux in~\cite{NKJ-M_tamingPN} hold to high orders (presumably to all orders) and contain the same exponential terms. Moreover, we can see where these factors arise in the calculation from a study of the structure of the MST solution used to compute $\Delta U$; see the Appendix. In particular, we can predict the coefficients of the first five appearances of a given power of $\log(R)$ in both the integer-order and half-integer terms. The only thing that prevents us from being able to predict further terms is the appearance of pieces that are not given by these simplifications at higher orders. 

Specifically, the higher-order logarithmic and half-integer terms in the full $\Delta U$ that the simplification predicts are given by the appropriate terms from
\begin{widetext}
\<\label{eq:ll_predictions}
\begin{split}\nonumber
&\quad\quad C^{\cS 1}_{22}\left[-\frac{128}{5}+\frac{5632}{105}\frac{1}{R}+\frac{452096}{6615}\frac{1}{R^2}-\frac{1654278784}{779625}\frac{1}{R^3}+\frac{436456134656}{178783605}\frac{1}{R^4}\right]\\
&\quad+ C^{\cS 1}_{21}\left[-\frac{32}{45}-\frac{80}{63}\frac{1}{R}-\frac{14234}{3969}\frac{1}{R^2}-\frac{61065898}{2338875}\frac{1}{R^3}\right] + C^{\cS 1}_{33}\left[-\frac{243}{7}+\frac{1215}{7}\frac{1}{R}-\frac{43254}{385}\frac{1}{R^2}-\frac{26690187609}{4904900}\frac{1}{R^3}\right]\\
&\quad+ C^{\cS 1}_{32}\left[-\frac{128}{63}+\frac{7424}{2835}\frac{1}{R}+\frac{2432}{1403325}\frac{1}{R^2}\right] + C^{\cS 1}_{31}\left[-\frac{1}{315}+\frac{1}{135}\frac{1}{R}-\frac{206}{10395}\frac{1}{R^2}-\frac{116962723}{1986484500}\frac{1}{R^3}\right]\\
&\quad + C^{\cS 1}_{44}\left[-\frac{32768}{567}+\frac{14024704}{31185}\frac{1}{R}-\frac{29572726784}{31216185}\frac{1}{R^2}\right] +  C^{\cS 1}_{43}\left[-\frac{729}{175} + \frac{6561}{385}\frac{1}{R}\right]\\
&\quad + C^{\cS 1}_{42}\left[-\frac{128}{3969}+\frac{34816}{218295}\frac{1}{R}-\frac{398729216}{1092566475}\frac{1}{R^2}\right] + C^{\cS 1}_{41}\left[-\frac{1}{11025} + \frac{103}{363825}\frac{1}{R}\right]\\
&\quad + C^{\cS 1}_{55}\left[-\frac{1953125}{19008} + \frac{798828125}{741312}\frac{1}{R}\right]  -\frac{524288}{66825}C^{\cS 1}_{54} + C^{\cS 1}_{53}\left[-\frac{2187}{17600} + \frac{19683}{20800}\frac{1}{R}\right]
-\frac{512}{200475}C^{\cS 1}_{52}\\
&\quad + C^{\cS 1}_{51}\left[-\frac{1}{4989600} + \frac{241}{194594400}\frac{1}{R}\right] -\frac{3359232}{17875}C^{\cS 1}_{66} - \frac{16777216}{47779875}C^{\cS 1}_{64} - \frac{512}{28667925}C^{\cS 1}_{62} - \frac{64}{5}C^{\cS 2}_{22}\\
&= %\{\text{logarithmic terms given in Eq.~\eqref{eq:DU_our}}\}\\
\{\text{leading logarithmic terms through $12.5$PN}\}\\
&\quad +\left[\left(\frac{28553817216903148928 }{3019990307709375}-\frac{40142209024}{17364375}\gamma   -\frac{80284418048 }{17364375 }\log(2)\right)\log^3(R) + \frac{5017776128}{17364375}\log^4(R)\right]\frac{1}{R^{14}}\\
&\quad + \frac{266856694218789586}{39313483295625}\frac{\pi\log^2(R)}{R^{14.5}} +\biggl[\bigg(\frac{12786084844378676616752}{335890033113009375}-\frac{230609748224  }{72930375}\gamma+\frac{613208455936 }{364651875}\log(2)\\
&\quad-\frac{19219356}{2401} \log (3)\biggr)\log^3(R) + \frac{28826218528 }{72930375 }\log ^4(R)\biggr]\frac{1}{R^{15}} + \biggl[\biggl(-\frac{116106125514835430463890813}{1773499374836689500000 }-\frac{80284418048 }{17364375}\pi^2\\
&\quad+\frac{8590432731136 }{607753125}\gamma+\frac{17180865462272}{607753125}\log(2)\biggr)\log^2(R)-\frac{4295216365568}{1823259375}\log^3(R)\biggr]\frac{\pi}{R^{15.5}}-\frac{740198388628949548 }{353821349660625}\frac{\log ^4(R)}{R^{16}}\\
&\quad + \biggl[\biggl(-\frac{5930646367149118394888262263}{29642775265127524500000}-\frac{5224670929064 }{364651875}\pi^2+\frac{21250374752872 }{1418090625 }\gamma-\frac{186725667394136}{12762815625}\log(2)\\
&\quad+\frac{749554884}{16807}\log(3)\biggr)\log^2(R)-\frac{10625187376436}{4254271875}\log^3(R)\biggr]\frac{\pi}{R^{16.5}} + \biggl[\biggl(\frac{2193470263750354274974196}{136035463410768796875}-\frac{8590432731136}{1823259375}\gamma\\
&\quad -\frac{17180865462272}{1823259375}\log(2)\biggr)\log^4(R)+\frac{4295216365568}{9116296875} \log ^5(R)\biggr]\frac{1}{R^{17}}+\frac{978084241468304898764}{58380522694003125}\frac{ \pi  \log
   ^3(R)}{R^{17.5}}
 \end{split}
 \?
 \<
 \begin{split}
 &\quad\hphantom{+\frac{8590432731136 }{607753125}\gamma+\frac{17180865462272}{607753125}\log(2)\biggr)\log^2(R)-\frac{4295216365568}{1823259375}\log^3(R)\biggr]\frac{\pi}{R^{15.5}}-\frac{740198388628949548 }{353821349660625}\frac{\log ^4(R)}{R^{16}}}\\
&\quad + \biggl[\left(\frac{4979471682011846405061446}{48032274735160340625}-\frac{475830406340432}{38288446875} \gamma -\frac{32617122056816 }{12762815625}\log(2) -\frac{374777442 }{16807}\log(3)\right)\log^4(R)\\
 &\quad+ \frac{237915203170216}{191442234375} \log ^5(R)\biggr]\frac{1}{R^{18}}
 + \biggl[ \biggl(-\frac{56225557382991466927743302446567}{359492757027834053373750000 }-\frac{68723461849088 }{5469778125}\pi^2\\
 &\quad+\frac{7353410417852416}{191442234375}\gamma+\frac{14706820835704832 }{191442234375 R^{37/2}}\log(2)\biggr)\log^3(R) -\frac{919176302231552 }{191442234375}\log^4(R)\biggr]\frac{1}{R^{18.5}}\\
&\quad + \{\text{$18$PN and higher leading logarithmic terms}\} + \{\text{other terms that are not accurate predictions}\},
\end{split}
\?
%\nkjm{Give higher orders, instead, as appropriate for the final version of $\Delta U$ we give?}
\end{widetext}
where
\begin{subequations}
\<
\begin{split}
C^{\cS 1}_{\ell m} &:= e^{2\bar{\nu}_{\ell m}\eulerlog_m(R)}\biggl[\frac{1}{2\bar{\nu}_{\ell m}}\frac{1}{R^2} - \frac{\bar{\nu}_{\ell m}}{4m}\frac{\pi}{R^{0.5}} - \frac{m\bar{\nu}_{\ell m}}{3}\frac{\pi^3}{R^{3.5}}\biggr]\\
&\quad\;\times\frac{1}{R^{\ell+1 + \varepsilon_{\ell m}}} - \frac{1}{2\bar{\nu}_{\ell m}}\frac{1}{R^{\ell + 3+ \varepsilon_{\ell m}}},
\end{split}
\?
\<
C^{\cS 2}_{\ell m} := \left[e^{2\bar{\nu}_{\ell m}\log(2/R)} - 1\right] \frac{1}{2\bar{\nu}_{\ell m}R^{5+2\ell+\epsilon_{\ell m}}},
\?
\end{subequations}
%\nkjm{Write all of these in terms of $u$ instead?}
and the ``appropriate terms'' that one should take from this expression are (as discussed above) the first five appearances of a given power of the logarithm in each of the integer and half-integer PN coefficients. (Note that the final subtracted terms in $C^{\cS 1}_{\ell m}$ and $C^{\cS 2}_{\ell m}$ are only necessary to remove some low-order nonlogarithmic integer-order terms, so one could leave off the subtracted term and simply ignore the additional terms, since they do not mix with the predictions of the simplification.) While this expression gives further appearances of these powers, such further appearances are \emph{not} predictions for complete coefficients of the full $\Delta U$, which obtains contributions from terms that are
not included in the simplification. (We give code that picks out the accurate predictions in the Supplemental Material~\cite{J-MSW_suppl}.) Moreover, even if there were no additional terms besides those given by the simplification, one would need to add on more terms in the series for a given mode, as well as additional modes, to obtain the sixth and higher appearances of a given power of a logarithm. Also, note that here we only need $\bar{\nu}_{\ell m}$ to second order [i.e., to $O(1/R^6)$; recall that $\nu$ is a series in $1/R^3$] in $C^{\cS 1}_{\ell m}$ and to first order [i.e., to $O(1/R^3)$]  in $C^{\cS 2}_{\ell m}$, and these higher-order terms are only needed for the highest few appearances of a given power of a logarithm.

It is also possible to obtain explicit expressions for the coefficients of the first few appearances of a given power of $\log(R)$ in the PN expansion of $\Delta U$ from Eq.~\eqref{eq:ll_predictions}, using the Taylor series for the exponential function. As
an illustration, we give here the expressions for the coefficients of the first two appearances of the $n$th power of $\log(R)$ in both the integer and half-integer PN terms of $\Delta U$:
\begin{widetext}
\<\label{eq:ll_preds_explicit}
\begin{split}
&\Biggl\{\frac{64}{5}\left(\frac{856}{105}\right)^{n-1}\frac{1}{R^2}  + \left[-\frac{2816}{105}\left(\frac{856}{105}\right)^{n-1} + \frac{16}{45}\left(\frac{214}{105}\right)^{n-1} + \frac{243}{14}\left(\frac{78}{7}\right)^{n-1} + \frac{1}{630}\left(\frac{26}{21}\right)^{n-1}\right]\frac{1}{R^3}\\
&- \frac{16}{5}\left(\frac{856}{105}\right)^{n+1}\frac{\pi}{R^{6.5}} + \left[\frac{704}{105}\left(\frac{856}{105}\right)^{n+1} - \frac{8}{45}\left(\frac{214}{105}\right)^{n+1} - \frac{243}{84}\left(\frac{78}{7}\right)^{n+1} - \frac{1}{1260}\left(\frac{26}{21}\right)^{n+1}\right]\frac{\pi}{R^{7.5}}\Biggr\}\frac{\log^n(R)}{n!R^{3n}}.
\end{split}
\?
\end{widetext}
Here one must only consider $n\geq 1$ in the integer-order terms, but can consider $n=0$ in the half-integer terms.
Note that the first appearance of a given power of $\log(R)$ (in either the integer or half-integer PN terms) comes solely from the leading term of the series multiplying $C^{S1}_{22}$ (i.e., just from the dominant $2,2$ mode), while the second appearance comes from the next term in the series multiplying $C^{S1}_{22}$ in addition to the leading terms multiplying $C^{S1}_{\ell m}$ for $(\ell, m) \in\{(2,1), (3,3), (3,1)\}$.

One can also similarly predict arbitrarily high-order ``leading logarithmic'' terms in the energy flux using the $S_{\ell m}$ factorization from~\cite{NKJ-M_tamingPN}, though this was not noted there: Here one obtains the coefficients of the first six occurrences of each power of a logarithm in the integer-order PN terms and the first five occurrences in the half-integer PN terms.  We give code that computes these predictions in the Supplemental Material~\cite{J-MSW_suppl}. Additionally, note that Nickel~\cite{Nickel} makes similar predictions of leading logarithmic-type terms to arbitrarily high orders for the ground state energy of $H_2^+$, and is able to derive some of them. Finally, these sorts of predictions of leading logarithms to arbitrarily high orders are likely related to the multipole moment beta functions discussed by Goldberger~\emph{et al.}~\cite{GR,GRR}, where they predict the coefficient of the first occurrence of a given power of a logarithm in both the energy flux and the binding energy using the beta function for the dominant $(2,2)$ mode. 

%%%%%%%%%%%%%%%%%
\section{Computing the infinite sum over renormalized $\ell$ modes to obtain the final result for $\Delta U$}
\label{sec:sum}
%%%%%%%%%%%%%%%%%

$\Delta U$, being a conservative invariant (and thus coming from the half-retarded plus half-advanced field, as discussed in Sec.~5 of~\cite{Barack-review}), requires a renormalization procedure where a noncontributing singular part of the retarded $\Delta U$ calculated at the position of the particle needs to be subtracted. This is done by using a mode-sum regularization technique where the retarded part is written as a sum over angular harmonics and the singular part is written as a sum over angular harmonics by extending it on the coordinate 2-sphere passing through the particle (i.e., at $r = r_0$). The explicit equations used in renormalization are given in~\cite{sf3}. The sum over $\ell$ modes converges quite slowly (the summand goes as $\ell^{-2}$), so it is customary to improve the convergence by finding
higher-order regularization coefficients numerically, as in~\cite{DMW,Detweiler,sf3}. However, to obtain an accuracy of $N$ digits in the final result of the renormalized $\Delta U$, one has to obtain these higher-order regularization coefficients to $N$ digits as well, which would necessitate going to prohibitively large $\ell$ (e.g., $\ell_\text{max} \sim 10^3$ for an accuracy of $5000$ digits). (The Wentzel-Kramers-Brillouin method detailed in~\cite{BD7} could be useful here in future work.)

Nevertheless, it is possible to obtain the analytic form for a given nonlogarithmic integer order PN term by obtaining the general form of the PN coefficient of a renormalized $\ell$ mode and performing the sum analytically, as in Bini and Damour~\cite{BD2,BD3,BD6}. We thus note that the $n$PN  coefficients of all the renormalized $\ell$ modes with $\ell\geq n-1$ are purely rational (with no transcendentals or logarithms). (Here we only consider $n\in\N$, since the half-integer terms only have a finite number of $\ell$ modes contributing.)  One can thus easily obtain analytic forms for these coefficients using PSLQ. (Here one scales with the denominator of the previous $\ell$ mode, to
help with the determination, and needs at most the values at four radii to obtain enough digits at $12$PN.) We then find that the general form of these coefficients as a function of $\ell$ can be expressed as linear combinations of members of a small family of
functions, namely
\begin{subequations}
\begin{align}
\cT_k^n(\ell) &:= \frac{1}{(\ell + k + 1/2)^n} +  \frac{(-1)^n}{(\ell - k + 1/2)^n},\\
\cU_k^n(\ell) &:= \frac{1}{(\ell + k)^n} +  \frac{(-1)^n}{(\ell - k + 1)^n},\\
\cV^n(\ell) &:= \frac{1}{(\ell + 1/2)^n}.
\end{align}
\end{subequations}
[Note that $\cV^n$ is the only one of these where the effect of the superscript $n$ is the same as taking $\cV$ to the $n$th power.]
These functions are similar to, though slightly more complicated than, the form considered for the regularization coefficients in Sec.~V of Shah~\emph{et al.}~\cite{sf3}.
We solve the linear system to obtain the coefficients (noting that one obtains excessively large rationals for the coefficients if one does not include the correct functions in the solve) and check that the expression successfully reproduces the values of the coefficients that were not used in the solve. We need to go to $\ell = 87$ (starting from $\ell = 16$, to avoid logarithmic terms at higher orders) for $12$PN, the most complicated case we consider, for a total of $72$ $\ell$-modes.

The general expressions for the first six PN coefficients have the form
\begin{gather}
\cT_1^1,\nonumber\\
\cT_{1-2}^1 \amp \cU_1^1,\nonumber\\
\cT_{1-3}^1 \amp \cT_1^2  \amp \cU_{1-2}^1  \amp \cV^2,\nonumber\\
\cT_{1-4}^1 \amp \cT_1^{2} \amp \cT_1^3 \amp \cU_{1-3}^1\amp \cV^2,\\
\cT_{1-5}^1 \amp \cT_{1-2}^2 \amp \cT_1^3 \amp \cU_{1-4}^1  \amp \cU_1^2 \amp \cU_1^3 \amp \cV^2,\nonumber\\
\cT_{1-6}^1 \amp \cT_{1-3}^2 \amp \cT_1^3 \amp \cT_1^4 \amp \cU_{1-5}^1 \amp \cU_{1-2}^2 \amp \cU_1^3 \amp \cV^2 \amp \cV^4,\nonumber
\end{gather}
where we just give the functions present, not the coefficients, and a range in a subscript indicates that all the functions in that range are present. We give the 
explicit expressions up to $12$PN in the electronic Supplemental Material~\cite{J-MSW_suppl}, and only note here that the specifics of the functions present grows in about the way one would expect: At
$n$PN, one has $\cT^k_{1-(n - 3k + 3)}$, $\cU^k_{1-(n - 3k + 2)}$ terms present, for $k$s with $n - 3k  + 3 \geq 1$ and $n - 3k  + 2\geq 1$, respectively, as well as $\cT^1_p$, $\cU^1_p$ terms for larger $p$ (where the specifics
of the terms present at a given PN order has a somewhat more complicated structure). One also has $\cV^k$ present for all even $k \leq (2/3)n$. For instance, at $12$PN, we have
\<
\begin{split}
&\quad \cT_{1-12}^1 \amp \cT_{1-9}^2 \amp \cT_{1-6}^3 \amp \cT_{1-3}^4 \amp \cT_1^5 \amp \cT_1^6  \amp \cT_1^7 \amp \cT_1^8 \\
& \amp \cU_{1-11}^1 \amp \cU_{1-8}^2  \amp \cU_{1-5}^3 \amp \cU_{1-2}^4 \amp \cU_1^5  \amp \cU_1^6  \amp \cU_1^7\\
& \amp \cV^2 \amp \cV^4 \amp \cV^6 \amp \cV^8.
\end{split}
\?
Also note that these general expressions diverge at the $\ell$s for which the PN coefficient is no longer purely rational, due to the $\cU^1_{n-1}$ term (cf.\ the discussion of the appearance of the logarithms at places where there are apparent poles in $\ell$ in the general form in Sec.~II~F of Bini and Damour~\cite{BD2}).

As Bini and Damour mention~\cite{BD2}, the infinite sums over these functions are straightforward to evaluate if one makes a partial fraction decomposition (and {\sc Mathematica}
will do them automatically without even needing to perform a partial fraction decomposition first): One finds that the sums over many of the terms telescope to a finite sum and the rest can be evaluated in terms of
the Riemann zeta function evaluated at even integers (giving even powers of $\pi$ with rational coefficients). Since these general expressions are not valid for the low-$\ell$ modes, where there
are also transcendentals present, one adds on the contributions from these low-order modes separately to obtain
the final expression. One finds that the size of the numerator and denominator of the final purely rational term is a good indicator of errors in the calculation: If one has
omitted a piece, or determined its analytic form incorrectly, this rational will be more complex than one would expect it to be, given the complexity at the previous order.

While we have analytic forms of the PN coefficients for $\Delta U$ through $12.5$PN, with the $13.5$PN term and all but the nonlogarithmic piece of the $13$PN also known, we only give the full $\Delta U$ to $11.5$PN here to save space. The analytic forms of these high-order coefficients are quite lengthy, even when written in eulerlog form. We give the previously known lower orders (which we have re-obtained) in their eulerlog form as well, for comparison, and to illustrate the structure. We also give the expression with the terms given by the simplifications removed, where
we go all the way to $12.5$PN. We give the full expressions for all these quantities in the electronic Supplemental Material~\cite{J-MSW_suppl}. Here we scale $\Delta U$ by $u := 1/R$ and write the expansion in terms of $u$, so that the coefficient of $u^n$ gives the $n$PN term of $\Delta U$. We also abuse notation (i.e., we use ``physicist's function definitions,'' not ``mathematician's function definitions'') and write $\eulerlog_m(u) := \gamma + \log(2mu^{1/2})$, which has the same value as the previous expression in terms of $R$ if one uses the $u$ related to this $R$, but is, of course, not given by substituting $u$ for $R$ in the previous expression.
%\nkjm{Do we even want to show everything here? We need to show/mention the $13.5$PN term, and the pieces of higher-order terms we also know, as well, of course...}
%\nkjm{Currently we use $R$ instead of $u$ as the variable in previous expressions. We should probably be consistent here...}
\begin{widetext}
\<\label{eq:DU_our}\nonumber
\begin{split}
\frac{\Delta U}{u} &= -1 -2u -5u^2 + \left[-\frac{121}{3}+\frac{41}{32}\pi^2\right]u^3 + \left[-\frac{1157}{15} + \frac{677}{512}\pi^2-\frac{128}{5}\eulerlog_2(u)\right]u^4 + \biggl[\frac{1606877}{3150}-\frac{60343}{768}\pi^2\\
&\quad - \frac{5}{7}\eulerlog_1(u) + \frac{5632}{105}\eulerlog_2(u) - \frac{243}{7}\eulerlog_3(u)\biggr]u^5 -\frac{13696}{525}\pi u^{5.5} +\biggl[\frac{17083661}{4050} - \frac{1246056911}{1769472}\pi^2\\
&\quad +\frac{2800873}{262144}\pi^4 -\frac{1193}{945} \eulerlog_1(u)+\frac{187904}{2835}\eulerlog_2(u)+\frac{1215}{7}\eulerlog_3(u) -\frac{32768}{567}\eulerlog_4(u)\biggr] u^6 \\
&\quad+ \frac{81077}{3675}\pi u^{6.5}+ \biggl[\frac{12624956532163}{382016250} - \frac{9041721471697}{2477260800}\pi^2 - \frac{23851025}{16777216}\pi^4 - \frac{2048}{5}\zeta(3) -\frac{1199567 }{332640}\eulerlog_1(u)  \\
&\quad -\frac{11564789888}{5457375}\eulerlog_2(u)- \frac{2873961}{24640}\eulerlog_3(u) +\frac{14024704 }{31185}\eulerlog_4(u) - \frac{1953125}{19008}\eulerlog_5(u)\\
&\quad +  \frac{109568}{525}\eulerlog_2^2(u) \biggr]u^7 + \frac{82561159}{467775}\pi u^{7.5}+ \biggl[-\frac{7516581717416867}{34763478750} - \frac{246847155756529}{18496880640}\pi^2\\
&\quad  + \frac{22759807747673}{6442450944}\pi^4 -\frac{41408}{105}\zeta(3) - \frac{64}{5}\log(2u) -\frac{31988738821 }{1222452000}\eulerlog_1(u) +\frac{80813099648}{33108075}\eulerlog_2(u) \\
&\quad  -\frac{85126268709}{15695680}\eulerlog_3(u)-\frac{67792273408}{70945875}\eulerlog_4(u)    + \frac{798828125}{741312}\eulerlog_5(u)-\frac{3359232}{17875}\eulerlog_6(u)\\
&\quad  + \frac{16022}{11025}\eulerlog_1^2(u) - \frac{4820992}{11025}\eulerlog_2^2(u) + \frac{18954}{49}\eulerlog_3^2(u) \biggr]u^8+ \biggl[-\frac{2207224641326123}{1048863816000} - \frac{219136}{1575}\pi^2 \\
&\quad+ \frac{23447552}{55125}\eulerlog_2(u)\biggr]\pi u^{8.5}+\biggl[-\frac{10480362137370508214933}{2044301131372500}-\frac{11665762236240841
}{226072985600}   \pi ^2\\
&\quad +\frac{32962327798317273 }{549755813888}\pi^4-\frac{27101981341 }{100663296} \pi ^6+ \frac{10221088}{2835}\zeta(3)- \frac{448}{5}\log(2u)- \frac{61470271483}{814968000}\eulerlog_1(u) \\
&\quad+ \frac{2840603616267776}{442489422375}\eulerlog_2(u) + \frac{8677864251603}{392392000}\eulerlog_3(u)- \frac{5946112890241024}{442489422375}\eulerlog_4(u) \\
&\quad - \frac{58533203125}{15567552}\eulerlog_5(u) + \frac{309049344}{125125}\eulerlog_6(u) - \frac{96889010407}{277992000}\eulerlog_7(u)+ \frac{51178}{19845}\eulerlog_1^2(u) \\
&\quad- \frac{5373212672}{9823275}\eulerlog_2^2(u) - \frac{94770}{49}\eulerlog_3^2(u) + \frac{1647312896}{1964655}\eulerlog_4^2(u)\biggr]u^9+\biggl[-\frac{30185191523470507}{12236744520000} \\
&\quad-\frac{1055996 }{11025}\pi^2+ \frac{1712534 }{1157625}\eulerlog_1(u) -\frac{1031692288}{1157625}\eulerlog_2(u) +\frac{246402}{343}\eulerlog_3(u)\biggr]\pi u^{9.5}
\end{split}
\?
\<
\begin{split}
\phantom{\frac{\Delta U}{u}}
&\quad+\biggl[-\frac{238946786344653264799175280203}{4522423405833558225000} -\frac{1070208441923650860489683 }{58656715985387520000}\pi^2+\frac{832229033014028790267991}{1662461581197312000} \pi^4\\
&\quad +\frac{54067065388369}{12884901888} \pi ^6-\frac{128695611256}{5457375} \zeta (3) +\frac{32768 }{5}\zeta (5) -\frac{20416}{35}\log(2u) -\frac{157982464536376957}{674943865596000} \eulerlog_1(u)\\
&\quad +\left(-\frac{1483437716511288604288}{14480466347221875}+ \frac{46895104}{33075}\pi^2 + \frac{3506176}{525} \zeta (3) \right)\eulerlog_2(u)-\frac{52813127885844357}{10492954472000}\eulerlog_3(u)\\
&\quad +\frac{8040008069311889408}{82745521984125}\eulerlog_4(u)-\frac{263296063591796875}{8742130068672}\eulerlog_5(u)-\frac{13640920722432
}{1146520375}\eulerlog_6(u)\\
&\quad+\frac{6491563697269}{1181466000}\eulerlog_7(u)-\frac{1099511627776
}{1688511825}\eulerlog_8(u)+\frac{54944178599}{7491884400}\eulerlog_1^2(u)\\
&\quad+\frac{69907855522816}{3781960875}\eulerlog_2^2(u)+\frac{79338802833 }{61661600}\eulerlog_3^2(u)-\frac{705049919488
}{108056025}\eulerlog_4^2(u)\\
&\quad+\frac{7548828125
}{4077216}\eulerlog_5^2(u)-\frac{187580416
}{165375} \eulerlog_2^3(u)\biggr]u^{10}+ \biggl[\frac{54441085537326639211}{3824681905044000} +\frac{78847804 }{66825}\pi^2\\
&\quad+ \frac{1096738}{416745}\eulerlog_1(u)-\frac{38224121176064}{34037647875}\eulerlog_2(u)-\frac{1232010}{343}\eulerlog_3(u)\\
&\quad+\frac{10351714238464}{6807529575}\eulerlog_4(u)\biggr]\pi u^{10.5}+ \biggl[\frac{3814229145040080910470246242071097}{13798711885916862654750000} \\
&\quad-\frac{497508986166915487823810257447 }{36601790774881812480000}\pi ^2+\frac{1213451006696869077146724173 }{2234348365129187328000}\pi^4-\frac{5276940898567193189 }{25975962206208}\pi ^6\\
&\quad-\frac{2759468242424 }{19864845} \zeta(3) +\frac{3283328}{105} \zeta (5) -\frac{175005952}{55125}\log(2u)+\frac{54784}{525}\log^2(2u)\\
&\quad+\left(-\frac{292720019838735815778069367}{313683671243842099200000}+\frac{1712534 }{694575}\pi ^2+\frac{128176 }{11025}\zeta (3)\right)
\eulerlog_1(u)\\
&\quad+\left(\frac{17152255889408499680050304}{325477442086506084375}-\frac{2063384576 }{694575}\pi ^2-\frac{154271744}{11025} \zeta(3)\right)\eulerlog_2(u)\\
&\quad+\left(-\frac{37422973611649093363871733}{81277585903754240000}+\frac{1232010}{343} \pi ^2 +\frac{1364688}{49} \zeta (3)\right) \eulerlog_3(u)\\
&\quad -\frac{1595162202161082218708992}{9299355488185888125}\eulerlog_4(u)+\frac{282725878023723294921875}{829173552753401856}\eulerlog_5(u)\\
&\quad-\frac{12192267599501090688}{202481230826875}\eulerlog_6(u)-\frac{60335405593501190803}{1792954994688000}\eulerlog_7(u)\\
&\quad+\frac{690493302243328}{57747104415}\eulerlog_8(u)-\frac{205891132094649}{168551219200}\eulerlog_9(u)+\frac{792734736884113}{14316991088400}\eulerlog_1^2(u)\\
&\quad-\frac{501413283015334912}{22370298575625}\eulerlog_2^2(u)+\frac{12234781198165473}{196392196000}\eulerlog_3^2(u)+\frac{1768982113681408}{127830277575}\eulerlog_4^2(u)\\
&\quad-\frac{3087470703125}{159011424}\eulerlog_5^2(u)+\frac{72640032768}{17892875}\eulerlog_6^2(u)
-\frac{6850136}{3472875}\eulerlog_1^3(u)+\frac{8253538304}{3472875}\eulerlog_2^3(u)\\
&\quad-\frac{985608}{343}\eulerlog_3^3(u) \biggr]u^{11}+\biggl[-\frac{45399846479271440442297518687}{663973981856472412125000}-\frac{10107325522351333 }{1311079770000}\pi^2+\frac{3506176}{23625} \pi ^4\\
&\quad+\frac{375160832 }{55125}\zeta (3)+\frac{839591622096533}{112490644266000} \eulerlog_1(u)+\left(\frac{2629370415206008832}{65522472159375} + \frac{375160832 }{165375}\pi ^2 \right)\eulerlog_2(u)\\
&\quad+\frac{732046976712531}{308616308000}\eulerlog_3(u)-\frac{4430533694062592 }{374414126625}\eulerlog_4(u)+\frac{5835244140625}{1749125664}\eulerlog_5(u)\\
&\quad-\frac{20071104512 }{5788125}\eulerlog_2^2(u)\biggr]\pi u^{11.5} + O(u^{12}).
\end{split}
\?
The terms through $13.5$PN for which we obtained analytic forms that we do not show here (i.e., all of these terms except for the nonlogarithmic $13$PN term) have the expected increase in
complexity, given the pattern at lower orders, and the complexity of the energy flux at infinity (see~\cite{Fujita22PN, NKJ-M_tamingPN}). In particular, we obtain a $\pi^8$ term
(from the sum over all $\ell$ modes) at $12$PN, along with an $\eulerlog_2(u)\log(2u)$ term [from the $(2,2)$ mode alone]. We also see the expected $\pi^4$ and $\zeta(5)$ terms in the linear
logarithmic term at $13$PN (which we obtain from our fit to the full $\Delta U$, as described below) and get the first $\log(2u)$ term in a half-integer piece at $13.5$PN.

If we write $\Delta U$ as a remainder plus the terms given by the two simplifications, we have, now going all the way to $12.5$PN,
\<
\begin{split}
\frac{\Delta U}{u} &= -1-2 u-5 u^2+\left[-\frac{121}{3} + \frac{41}{32} \pi ^2\right] v^3+\left[-\frac{1157}{15} + \frac{677 }{512}\pi ^2\right]u^4+\left[\frac{1606877}{3150}-\frac{60343}{768}\pi^2\right]u^5\\
&\quad +\left[\frac{17083661}{4050}-\frac{1246056911}{1769472} \pi ^2+\frac{2800873}{262144} \pi
   ^4\right] u^6+\left[\frac{12624956532163}{382016250}-\frac{8826302018257 }{2477260800}\pi ^2-\frac{23851025 }{16777216}\pi ^4\right]u^7\\
   &\quad +\left[-\frac{7516581717416867}{34763478750} -\frac{741674600438227 }{55490641920}\pi^2+\frac{22759807747673}{6442450944} \pi ^4\right]u^8\\
   &\quad +\biggl[-\frac{10480362137370508214933}{2044301131372500}-\frac{25850880135908623907}{494421619507200} \pi^2+\frac{32962327798317273}{549755813888} \pi ^4-\frac{27101981341}{100663296} \pi^6\biggr]u^9\\
   &\quad+\biggl[-\frac{238946786344653264799175280203}{4522423405833558225000}-\frac{18927900985563784496607041 }{1583731331605463040000}\pi^2\\
   &\quad +\frac{166464310992954697947563 }{332492316239462400}\pi ^4+\frac{54067065388369 }{12884901888}\pi^6\biggr]u^{10} -\frac{8192 }{75}\pi  u^{10.5}\\
   &\quad+\biggl[\frac{3814229145040080910470246242071097}{13798711885916862654750000}-\frac{31306151918920018820376206813081 }{2305912818817554186240000}\pi^2\\
   &\quad+\frac{48548370032386602537826133}{89373934605167493120} \pi ^4-\frac{5276940898567193189 }{25975962206208}\pi^6\biggr]u^{11}-\frac{2192896 }{6741}\pi  u^{11.5}\\
   &\quad+   \biggl[\frac{176538264526096039025674251863176559931511}{12080725340499801136900598850000}-\frac{230595271714866856972896270561913344391 }{558842583466071892143636480000}\pi^2\\
   &\quad-\frac{902366950567138330511081959572149}{18875774988611374546944000} \pi^4 -\frac{167517791710253563186933}{26599385299156992} \pi ^6+\frac{44336492264184971}{2473901162496} \pi ^8\\
   &\quad+ \frac{65536}{75}\log(2u)\eulerlog_2(u)\biggr]u^{12}-\frac{31486592}{43335} \pi u^{12.5} + \sum_{\ell = 2}^{10}\sum_{m = 1}^\ell C^{[1]}_{\ell m}\Upsilon^{\cS 1}_{\ell m} + \sum_{\ell = 2}^{4}\sum_{m = 1}^\ell C^{[2]}_{\ell m}\Upsilon^{\cS 2}_{\ell m} + O(u^{13}),
\end{split}
\?
\end{widetext}
where $\Upsilon^{\cS 1}_{\ell m}$ and $\Upsilon^{\cS2}_{\ell m}$ are integer order power series in $u$ with rational coefficients, which we give (to the order known) in the electronic Supplemental Material~\cite{J-MSW_suppl} [see Eqs.~\eqref{eq:A22} and~\eqref{eq:B22} for the expressions for the $(2,2)$ mode of $\Delta U/U$]. (Note that the odd $m$ terms for the $\ell = 10$ $\Upsilon^{\cS 1}_{\ell m}$s and the $\ell = 4$ $\Upsilon^{\cS2}_{\ell m}$s do not contribute until $13$PN.)

We find that the $13.5$PN piece of $\Delta U$ has more terms that are not removed by the simplification than do the previous half-integer PN terms, just as occurs at this order in the energy flux (see the expression for the $S_{\ell m}$ factorisation of $\eta_{22}$ in the electronic Supplemental Material for~\cite{NKJ-M_tamingPN}), and, as in the energy flux, the additional terms
all come from the dominant $(2,2)$ mode at this order. Specifically, the $13.5$PN piece of $\Delta U/u$ remaining after subtracting off the parts given by the simplification is
\<
\begin{split}
&\biggl[-\frac{2096793662144}{139033125}-\frac{131072 }{225}\pi ^2 + \frac{14024704 }{7875}\eulerlog_2(u)\\
& +\frac{7012352}{2625}\log(2u)\biggr]\pi u^{13.5}.
\end{split}
\?
However, the portion remaining in other PN coefficients of $\Delta U$ after using the simplification does not have exactly the same structure as that in $\eta_{22}/|S_{22}|^2$. For instance, $\eta_{22}/|S_{22}|^2$ also has $\eulerlog_2$ and $\eulerlog_2^2$ terms in the $12$PN coefficient.

%----------------------
\subsection{Checking the results for $\Delta U$ by making an independent fit}
\label{sec:fit}
%----------------------

We performed an independent check of these results by making a fit for the PN coefficients of $\Delta U$ using data at smaller radii and the fit procedure described in SFW~\cite{SFW}. In addition to checking the decimal expansions of the terms we have already obtained analytically, we also implicitly check all the coefficients we have obtained in the fit by verifying that the higher-order coefficients are not too large, as described below. We perform these fits iteratively, obtaining analytic forms for
as many terms as possible with the accuracy obtained from a given fit, subtracting these off, and fitting again. In this case, we proceeded through six iterations, where the first fit only went to $20$PN, and these coefficients were obtained with an accuracy of just a few digits, while at the fifth and final iteration, after we had subtracted off $48$ coefficients, we obtained the $20$PN coefficients that we did not obtain analytically to $\sim 41$ digits, and were
able to go all the way to $21.5$PN, where we obtained the coefficients we did not know analytically from the simplification to $\sim 10$ digits.

We made verifications of these results by checking that the analytic forms we obtain have the expected forms, and that the terms given by the simplification agree, in addition to the stringent verification provided by the fit itself, described below.
We used the simplification to aid this process, so we needed to include at most $3$ transcendentals in the vector to which we apply PSLQ (for the $16.5$PN linear logarithmic term). This procedure (of using PSLQ to iteratively improve a fit, aided by a conjecture for the form of certain leading logarithm-type terms) is very similar to the one used by Nickel to obtain high-order terms in the expansion of the ground state energy of $H_2^+$ in~\cite{Nickel}.

We give the final results of this fit (both analytical and numerical) in the electronic Supplemental Material~\cite{J-MSW_suppl}, including showing the remainder of the analytic terms after removing the portions given by the simplification.

One advantage of using high-precision data to extract PN coefficients is that it is relatively easy to check the accuracy of the analytical coefficients calculated using PSLQ. If we had used an incorrect coefficient, say for an $n$PN nonlogarithmic term, and used it to find other coefficients, the coefficient of the $n$PN higher logarithmic terms and subsequent higher order PN coefficients would have increased by many orders of magnitude to give a nonsensical result.

Let us illustrate this with an example. The numerically extracted $21.5$PN nonlogarithmic coefficient ($\alpha_{21.5}$, in the terminology of SFW) has a size of $\sim 10^{11}$. If we had used an incorrect $21$PN $\log^5(R)$ term ($\zeta_{21}$), the $\alpha_{21.5}$ we extracted from the fit would have increased to a size of about $10^{40}$, a nonsensical result. The analytical form of $\zeta_{21}$ was determined from its numerical value, which was extracted with an accuracy of $13$ significant digits. So, to test the sensitivity of the fit to the values of the digits we did not extract, we inject random analytical (absolute) errors in $\zeta_{21}$ of magnitude ranging from $10^{-13}$ to $10^{-27}$ and extract $\alpha_{21.5}$. These errors are injected by using random numbers between $1000$ and $5000$, multiplied with powers of 10 ranging from $-16$ to $-30$. We see that if we had included an error of magnitude $10^{-13}$, the numerically extracted $\alpha_{21.5}$ would have had a size of $\sim 10^{47}$, and if we had included an error of magnitude $10^{-27}$ (which is more than twice the number of significant digits used to calculate the analytical form of $\zeta_{21}$), $\alpha_{21.5}$ would have had a size of $\sim 10^{35}$.

This example clearly demonstrates the sensitivity of the numerical fitting technique we use and how the analytical forms of numerically extracted PN coefficients can be checked by injecting errors. 
Of course, it is always possible to have a quantity that only differs from a reasonable-looking analytic form at extremely high positions in its decimal expansion (see some of the examples given by Bailey and Borwein~\cite{BB_AMS,BB_AMS2}). However, this seems quite unlikely to be the case here, particularly because we have a good idea of the form of the coefficients and the growth of their complexity, from the forms of lower orders and the PN expansion of the energy flux at infinity.

%%%%%%%%%%%%%%%%%
\section{Convergence}
\label{sec:conv}
%%%%%%%%%%%%%%%%%

\begin{figure*}[htb]
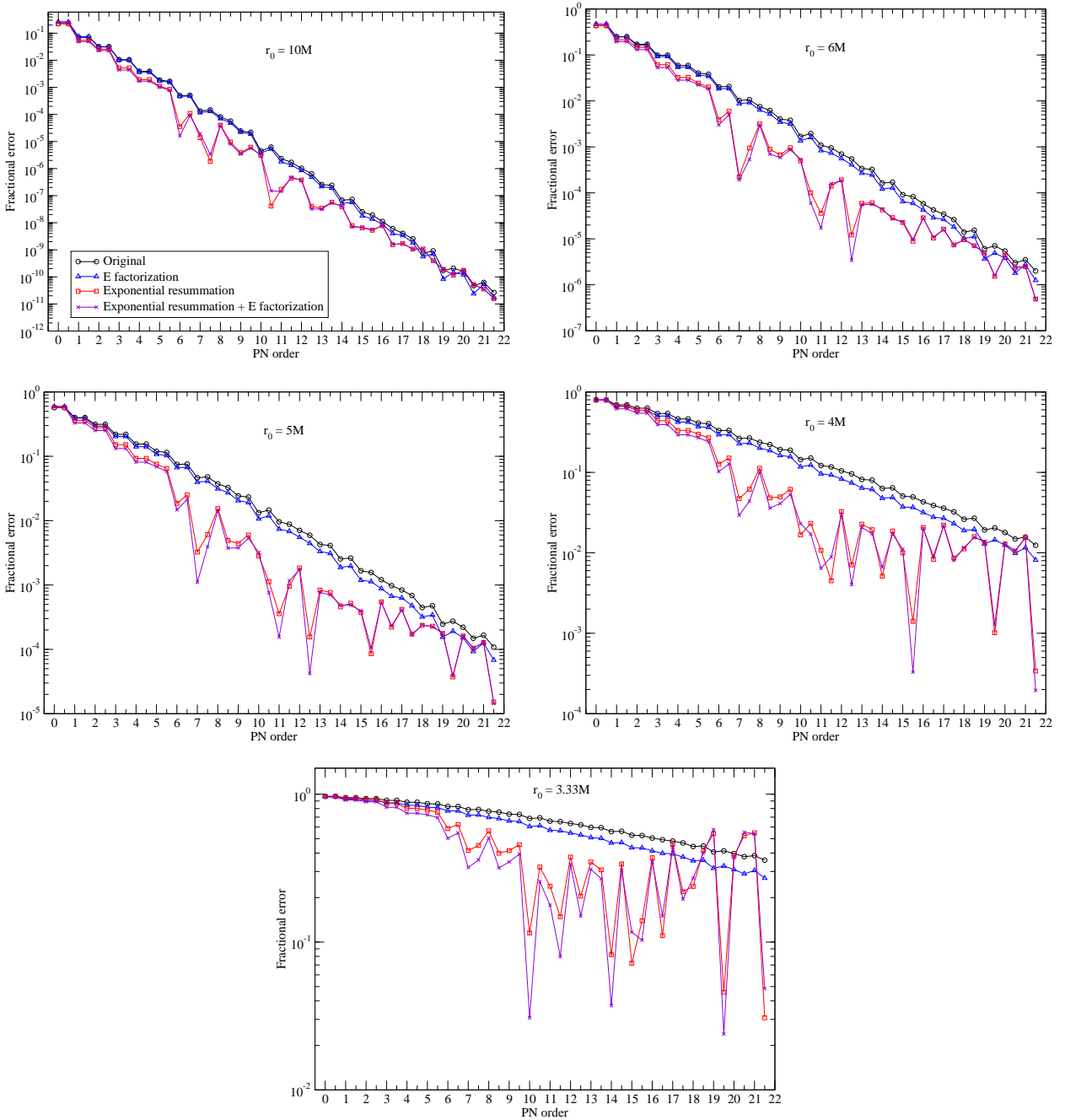

\centering
\subfloat{
\epsfig{file=DeltaU_conv_comp_r10_215PN.eps,width=8cm,clip=true}
}
\quad
\subfloat{
\epsfig{file=DeltaU_conv_comp_r6_215PN.eps,width=8cm,clip=true}
}\\
\subfloat{
\epsfig{file=DeltaU_conv_comp_r5_215PN.eps,width=8cm,clip=true}
}
\quad
\subfloat{
\epsfig{file=DeltaU_conv_comp_r4_215PN.eps,width=8cm,clip=true}
}\\
\subfloat{
\epsfig{file=DeltaU_conv_comp_r333_215PN.eps,width=8cm,clip=true}
}
\caption{\label{fig:DU_conv} Convergence of the $21.5$PN expression for $\Delta U$ for orbits at various radii, comparing with the numerical data from Dolan~\emph{et al.}~\cite{Dolanetal-tidal} and Akcay~\emph{et al.}~\cite{ABDS}. Specifically, we show the convergence of the plain series, as well as the results of factoring out the test particle binding energy and/or performing exponential resummation on the entire series.}
\end{figure*}

\begin{figure}[htb]
\epsfig{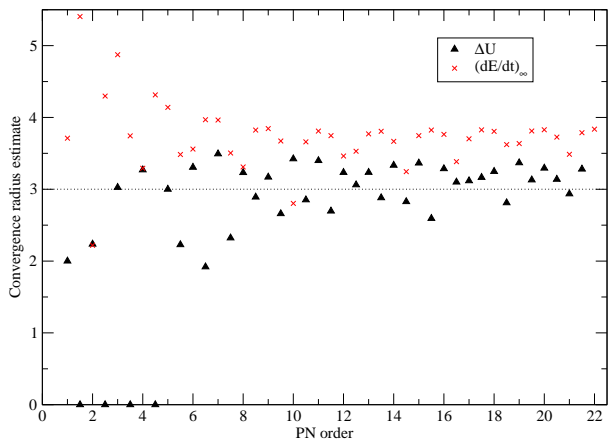}
\caption{\label{fig:DU_conv_radius_comp} An estimate of the Schwarzschild radial coordinate (in $M$) of the radius of convergence of the PN series for $\Delta U/u$, obtained from $a_n^{2/n}$, where $a_n$ denotes the nonlogarithmic coefficient of $v^n$. We also show the same estimate for the test particle energy flux at infinity $(dE/dt)_\infty$, scaled by the Newtonian energy flux (from~\cite{Fujita22PN}), for comparison.}
\end{figure}

It is interesting to consider the convergence of the high-order PN expression we have obtained for $\Delta U$. In Fig.~\ref{fig:DU_conv}, we compare the convergence of the plain $21.5$PN expansion of $\Delta U$ with various resummations. Here we compare with the numerical data from Table~III in Dolan~\emph{et al.}~\cite{Dolanetal-tidal} for radii of $\{4,5,6,10\}M$ and with data from Table~IX in Akcay~\emph{et al.}~\cite{ABDS} for a radius of $(10/3)M \simeq 3.33M$, converting their $h_{uu}^{R,L}(x)$ into our $\Delta U$ using their Eq.~(17) and Eq.~(2) in~\cite{SFW}. We find that while the rate of convergence decreases as the radius of the orbit decreases (as expected), the series still converges reasonably well inside the innermost stable circular orbit (ISCO) at $r = 6M$, and continues to converge quite monotonically close to the light ring at $r= 3M$, albeit extremely slowly. Moreover, the exponential resummation (of the entire series, as originally proposed by Isoyama~\emph{et al.}~\cite{Isoyamaetal}, not mode-by-mode, as in~\cite{NKJ-M_tamingPN,Fujita11PNKerr}) improves the convergence substantially for low to medium orders, particularly within the ISCO, though it makes it significantly less monotonic, and actually worsens the convergence at high orders in the strong field regime.

If one performs a partial mode-by-mode exponential resummation, either exponentially resumming the modes through $\ell = 10$ and the remainder of the full $\Delta U$ separately, or using the simplifications on the modes and exponentially resumming the portions that multiply the simplifications, as well as the remainders of the modes, this does not perform better than exponential resummation applied to the entire expression (though it also does not  behave as erratically as full exponential resummation at high orders in the strong field). If one just applies exponential
resummation to the individual modes, then one finds that it does improve the convergence of some modes, particularly the ones with larger $\ell -m$.
Factoring out the test particle binding energy, as done in Akcay~\emph{et al.}~\cite{ABDS} also improves the convergence, particularly near the light ring (where the test particle binding energy diverges), but does not improve the convergence nearly as much as the exponential resummation on its own.

One can also estimate the radius of convergence (in $v$) of the PN series for $\Delta U$ by looking at $a_n^{-1/n}$, where $a_n$ is the nonlogarithmic coefficient of $v^n$. If the series has no logarithmic terms, then the radius of convergence, $v_r$, is given by $1/v_r = \limsup_{n\to\infty}a_n^{1/n}$. Thus, since $\Delta U$ diverges at the light ring (as discussed in, e.g.,~\cite{ABDS}), one expects the radius of convergence of the PN series for $\Delta U$ one estimates in this way by taking only the nonlogarithmic portion of the coefficients (and just considering the known orders, without taking the limit) to be close to the light ring. Or, to put it another way, one expects the size of the coefficients of $v^n$ to grow approximately like $3^{n/2}$. One indeed sees that this is the case, as shown in Fig.~\ref{fig:DU_conv_radius_comp}, where we plot the Schwarzschild radial coordinate of the radius of convergence estimated in this fashion. We also plot the growth of the PN
coefficients of the energy flux at infinity, for comparison.

%%%%%%%%%%%%%%%%%
\section{Conclusions and outlook}
\label{sec:concl}
%%%%%%%%%%%%%%%%%

We have introduced a method for obtaining analytic forms of high-order post-Newtonian coefficients to linear order in the mass ratio from high-accuracy numerical results from black hole perturbation theory. We have also given  the first application of this method to the case of Detweiler's redshift invariant, which (when evaluated in linear black hole perturbation theory) gives the linear in mass ratio piece of the binary's binding energy and the EOB radial potential. Here we have found analytic forms for all these coefficients to $12.5$PN, and have obtained mixed analytic-numerical results to $21.5$PN (including analytic forms for the complete $13.5$PN term, and all but the nonlogarithmic piece of the $13$PN term), substantially improving on the previous $9.5$PN knowledge of this quantity. We also found a simplification of the individual modes, similar to that found for the energy flux at infinity in~\cite{NKJ-M_tamingPN}, which also allows us to predict certain leading logarithmic-type terms to all orders in the full $\Delta U$.

The new terms we have obtained improve the accuracy of the series, even inside the ISCO and near the light ring (though the convergence there is very slow, as expected); factoring out the energy, which diverges at the light ring, improves the convergence somewhat. Since exponential resummation of the individual modes of radiative quantities improves the convergence much more than exponential resummation of the full quantity (see~\cite{NKJ-M_tamingPN,Fujita11PNKerr}), we had hoped that there might be a better way of performing the exponential resummation here, which would behave better in the strong-field regime. However, our experiments in this regard were unsuccessful, in that we only obtained very modest improvements, much less than the best improvement of exponential resummation applied to the full series, though the improvements did not have the full exponential resummation's erratic behavior.

It might also be possible to use these high-order perturbative results to improve convergence by finding nonperturbative pieces, using resurgence (see, e.g.,~\cite{DU} for an application of these ideas in quantum mechanics). Another possibility would be to try to resum the purely integer-order PN series with rational coefficients that enter into the simplification or its remainder, as was
done for a (likely considerably simpler) self-force series in~\cite{BPN}.

We are now in a position to apply this method to the much more difficult case of perturbations of the Kerr metric. Here we will likely combine a study of $\Delta U$ with a study of the structure of the energy flux at infinity (computed numerically to $20$PN in~\cite{SKerr} and analytically to $11$PN in~\cite{Fujita11PNKerr}), since our previous study of this structure in the Schwarzschild case~\cite{NKJ-M_tamingPN} was very useful in the present calculation.

%%%%%%%%%%%%%%%%%
\acknowledgments
%%%%%%%%%%%%%%%%%

We thank Marc Casals, John L.\ Friedman, and Adam Pound for useful discussions, and an anonymous referee for useful comments and suggestions. We also thank Chris Kavanagh, Adrian Ottewill, and Barry Wardell for sharing their results for comparison. NKJ-M acknowledges support from the DFG SFB/TR7 and the AIRBUS Group Corporate Foundation
through a chair in ``Mathematics of Complex Systems'' at the
International Centre for Theoretical Sciences. AGS was supported by the European Research Council under the European Union's Seventh Framework Programme (FP7/2007-2013)/ERC grant agreement no.\ 304978.
BFW acknowledges support from NSF Grant PHY 1205906, and hospitality from the University of Southampton at an early stage in this work.

%############
\appendix*
%############

%%%%%%%%%%%%%%%%%%%%%%%%%
\section{Obtaining the $e^{2\bar{\nu}_{\ell m}\eulerlog_m(R)}$ and $e^{2\bar{\nu}_{\ell m}\log(2/R)}$ contributions to the simplifications of the modes of $\Delta U$}
\label{app:exp_factor}
%%%%%%%%%%%%%%%%%%%%%%%%%

Just as one can see where the $S_{\ell m}$ and $V_{\ell m}$ factorizations of the energy flux from~\cite{NKJ-M_tamingPN} arise from the MST formalism (as discussed in Sec.~IV of~\cite{NKJ-M_tamingPN}), it should be possible to see how the simplifications for $\Delta U$ we have found [Eqs.~\eqref{eq:Upsilon_simp} and~\eqref{eq:Upsilon_simpC2}] 
arise from the MST formalism, and (in the best case) predict higher-order terms in them. However, we shall see that the situation for $\Delta U$ is more complicated than that for the energy flux, and will at present content
ourselves with seeing how the $e^{2\bar{\nu}_{\ell m}\eulerlog_m(R)}$ and $e^{2\bar{\nu}_{\ell m}\log(2/R)}$ contributions to the simplifications arise. Note that here we shall expand in $v$ instead of $R$, for simplicity, and to avoid confusion with some other quantities named $R$.

Specifically, if one looks at Eq. (29) in~\cite{sf4} and our Eqs.~\eqref{eq:psi0} and~\eqref{eq:Psi}, one finds that the modes of $\Delta U$ have the form
\<\label{Upsilon_sim}
\Upsilon_{\ell m} \sim \frac{R^\text{in}R^\text{up}}{W[R^\text{in},R^\text{up}]} + \text{c.c.},
\?
where we have noted that $\Delta U$ comes from the metric perturbation and are using the same notation as in Sec.~IV of~\cite{NKJ-M_tamingPN}, where $\sim$ denotes that we are neglecting any terms that do not lead to
transcendentals and logarithms (including the overall scaling). We have suppressed the dependence of $R^\text{in}$ and $R^\text{up}$ on $\ell$ and $m$ here (and in similar expressions later), for simplicity. Note that
in the expressions in previous sections we denote $R^\text{in}$ and $R^\text{up}$ by $R_H$ and $R_\infty$, respectively.
Also, we have (Eq.~(166) in Sasaki and Tagoshi~\cite{ST})
\<
R^\text{in} = K_\nu R_\text{C}^\nu + K_{-\nu - 1}R^{-\nu - 1}_\text{C}
\?
($K_\nu$ and $R_\text{C}^\nu$ are given in, e.g., Eqs.~(6) and~(7) in~\cite{NKJ-M_tamingPN}) and [Eqs.~(4.1) and~(4.9) in~\cite{MST}, evaluated for $|s| = 2$]
\<
R^\text{up} = \frac{S_\nu R_\text{C}^\nu - \ri e^{\ri\pi\nu}R^{-\nu - 1}_\text{C}}{S_\nu + e^{2\ri\pi\nu}},
\?
where we have defined
\<
S_\nu := \frac{\sin\pi(\nu + \ri\epsilon)}{\sin\pi(\nu - \ri\epsilon)},
\?
and (Eq.~(23) in Sasaki and Tagoshi~\cite{ST})
\<
W[R^\text{in},R^\text{up}] 
\sim  C^\text{trans}B^\text{inc}
\?
denotes the Wronskian of $R^\text{in}$ and $R^\text{up}$.
%(Note that~\cite{SFW}  and~\cite{sf4} denote $R^\text{in}$ and $R^\text{up}$ by $R_H$ and $R_\infty$, respectively.)
Here [Eqs.~(157), (158), (168), and (170) in Sasaki and Tagoshi~\cite{ST}, noting that $\kappa = 1$ for Schwarzschild]
\begin{subequations}
\begin{gather}
C^\text{trans} 
\sim A_-^\nu \epsilon^{\ri\epsilon},\\
B^\text{inc} \sim \left(K_\nu - \ri e^{-\ri\pi\nu}S_\nu K_{-\nu-1}\right)A_+^\nu\epsilon^{-\ri\epsilon},\\
A_+^\nu \sim 2^{-\ri\epsilon}e^{-\pi\epsilon/2}e^{\ri\pi\nu/2}\frac{\Gamma(1 + \nu + \ri\epsilon)}{\Gamma(1 + \nu - \ri\epsilon)},\\
A_-^\nu \sim 2^{\ri\epsilon}e^{-\pi\epsilon/2}e^{-\ri\pi\nu/2}.
\end{gather}
\end{subequations}

We thus have
\<
\begin{split}
W[R^\text{in},R^\text{up}] &\sim \left(K_\nu - \ri e^{-\ri\pi\nu}S_\nu K_{-\nu-1}\right)A_+^\nu A_-^\nu\\
&\sim \left(K_\nu - \ri e^{-\ri\pi\nu}S_\nu K_{-\nu-1}\right)\frac{\Gamma(1 + \nu + \ri\epsilon)}{\Gamma(1 + \nu - \ri\epsilon)}e^{-\pi\epsilon},
\end{split}
\?
so we can write Eq.~\eqref{Upsilon_sim} as
\<
\begin{split}
\Upsilon_{\ell m} &\sim e^{\pi\epsilon} \frac{\Gamma(1 + \nu - \ri\epsilon)}{\Gamma(1 + \nu + \ri\epsilon)}\frac{R_\text{C}^\nu\left(S_\nu R_\text{C}^\nu - \ri e^{\ri\pi\nu}R^{-\nu - 1}_\text{C}\right)}{S_\nu + e^{2\ri\pi\nu}}\\
&\quad\times\frac{1 + \frac{K_{-\nu - 1}}{K_\nu}\frac{R^{-\nu - 1}_\text{C}}{R_\text{C}^\nu}}{1 - \ri e^{-\ri\pi\nu}S_\nu \frac{K_{-\nu-1}}{K_\nu}} + \text{c.c. }.
\end{split}
\?
The $K_{-\nu-1}R^{-\nu - 1}_\text{C}/(K_\nu R_\text{C}^\nu)$ term is likely the origin of the $e^{2\bar{\nu}_{\ell m}\log(2/R)}$ contribution to the $\Upsilon^{C2}_{\ell m}$ simplification (just as it is for the $V_{\ell m}$ simplification in~\cite{NKJ-M_tamingPN}), since $K_{-\nu-1}R^{-\nu - 1}_\text{C}/(K_\nu R_\text{C}^\nu) \sim (2v^2)^{2\nu}\{\text{gamma function terms}\}$ (cf.\ Eqs.~(27a) and~(27c) in~\cite{NKJ-M_tamingPN}). As these $K_{-\nu-1}/K_\nu$ terms only contribute at higher orders, as discussed in Sec.~IV of~\cite{NKJ-M_tamingPN}, we shall thus omit the
final fraction in the product in the ensuing discussion, where we are concerned with the $\Upsilon^{C1}_{\ell m}$ simplification.

Now (recalling that $\epsilon = 2mv^3$ and $\omega r_0 = mv$), we have
\<
\label{RC}
R_\text{C}^\nu \sim \left(1 - 2v^2\right)^{-2\ri mv^3}e^{-\ri mv}(2mv)^\nu \frac{\Gamma(1 + \nu + \ri\epsilon)}{\Gamma(1+ 2\nu)}.
\?
Thus, the $R_\text{C}^\nu R^{-\nu - 1}_\text{C}$ term in $\Upsilon_{\ell m}$ contributes
\<
%\begin{split}
\sim e^{\pi\epsilon}\frac{\Gamma(1 + \nu - \ri\epsilon)\Gamma(1 - \nu + \ri\epsilon)}{\Gamma(1+ 2\nu)\Gamma(1- 2\nu)}\cX,
%\end{split}
\?
where
\<
\cX := \frac{\left(1 - 2v^2\right)^{-4\ri mv^3}e^{\ri(\pi\nu -2 mv)}}{S_\nu + e^{2\ri\pi\nu}} + \text{c.c.}\;.
\?
This cannot contribute any eulerlog terms (the expansion of the gamma functions does not contain a $\gamma$), so we leave it alone.

The $\left(R_\text{C}^\nu\right)^2$ term in $\Upsilon_{\ell m}$, on the other hand, does give exactly the eulerlog contribution found in $\Upsilon^{C1}_{\ell m}$. Specifically, it gives
\<
\begin{split}
&\sim e^{\pi\epsilon}(2mv)^{2\nu}\frac{\left|\Gamma(1 + \nu + \ri\epsilon)\right|^2}{\left[\Gamma(1+ 2\nu)\right]^2}\tilde{\cX}\\
&= \tilde{\cX}\exp\left[2\nu\eulerlog_m(v) + 2\pi mv^3+ \sum_{n=2}^\infty\frac{\zeta(n)}{n}\cG\right].
\end{split}
\?
where
\<
\begin{split}
\tilde{\cX} &:= \frac{\left(1 - 2v^2\right)^{-4\ri mv^3}e^{-2\ri m v}}{1 + e^{2\ri\pi\nu}/S_\nu} + \text{c.c.},\\
\cG &:= (-\nu -2\ri mv^3)^n + (-\nu +2\ri mv^3)^n - 2(-2\nu)^n.
\end{split}
\?
[Here we have abused notation in the ``physicist's way'' with $\eulerlog_m$ again, writing $\eulerlog_m(v) := \gamma + \log(2mv)$, which is not what one would obtain when substituting $v$ for the
argument of either of the previous two definitions, but, of course, agrees with them when one substitutes the values of $R$ or $u$ corresponding to this $v$.]
Unfortunately, the process of obtaining the full simplification from a study of the pieces entering the MST computation is obviously more subtle in this case than
it is for the energy flux (discussed in~\cite{NKJ-M_tamingPN}): The remaining terms in the expansion of this quantity [i.e., leaving off the $e^{2\nu\eulerlog_m(v)}$ factor] are not
those found from a study of the expansion of $\Upsilon_{\ell m}$ and given in Eq.~\eqref{eq:Upsilon_simp}. The terms obtained from this expansion are more numerous and do not have the correct coefficients. The leading term indeed has the factor of $1/\nu$, but none of the other terms seem to match.

 \bibliography{SFtoexactPN}

\end{document}